\documentclass[useAMS,usenatbib]{mn2e}

\usepackage{amsmath}
\usepackage{amssymb}
\usepackage[parfill]{parskip}   
\usepackage{ifpdf}
\usepackage{graphicx} 
\usepackage{longtable}
\usepackage{rotating}
\usepackage{epsf}
\usepackage{bm}
\usepackage{subfigure}
\usepackage{morefloats}
\usepackage{gensymb}
\usepackage{verbatim}
\usepackage{array}
\usepackage{ragged2e} 
\usepackage{color}
\usepackage{soul}
\usepackage{lscape}

\newcommand{\Halpha}{{H$\alpha$}}
\newcommand{\HI}{{\sc Hi}}
\newcommand{\HII}{{\sc Hii}}
\newcommand{\HIPASS}{{\sc HiPASS}}
\newcommand{\kms}{\mbox{km\thinspace s$^{-1}$}}

\graphicspath{{figures/}}

\begin{document} 

\title{Choirs, HI Galaxy Groups: \\Catalogue and Detection of Star-forming Dwarf Group Members}
\author[S. Sweet et al.]{Sarah M. Sweet$^{1}$, Gerhardt Meurer$^{2,3}$, Michael J. Drinkwater$^{1}$, Virginia Kilborn$^{4}$, 
\newauthor Helga D\'{e}nes$^{4}$, Kenji Bekki$^{2,3}$, Dan Hanish$^{5}$, Henry Ferguson$^{6}$, Patricia Knezek$^{7}$,
\newauthor Joss Bland-Hawthorn$^{8}$, Michael Dopita$^{9}$, Marianne T. Doyle-Pegg$^{1}$, Ed Elson$^{2,3}$, 
\newauthor Ken Freeman$^{9}$, Tim Heckman$^{5}$,  Robert Kennicutt$^{10,11}$, Ji Hoon Kim$^{12}$, B\"{a}rbel Koribalski$^{13}$, 
\newauthor Martin Meyer$^{2,3}$,  Mary Putman$^{14}$, Emma Ryan-Weber$^{10}$, Chris Smith$^{15}$,  
\newauthor Lister Staveley-Smith$^{2,3}$, O. Ivy Wong$^{13}$, Rachel Webster$^{16}$, Jessica Werk$^{14}$, Martin Zwaan$^{17}$\\
1: School of Mathematics and Physics, University of Queensland, Qld, 4072, Australia; sarah@sarahsweet.com.au;\\
2: School of Physics, University of Western Australia, 35 Stirling Highway, Crawley, WA, 6009, Australia; \\
3: International Centre for Radio Astronomy Research, ICRAR M468, 35 Stirling Highway, Crawley, WA, 6009, Australia; \\
4: Mail number H30, Swinburne University of Technology, PO Box 218, Hawthorn, Victoria 3122, Australia;\\
5: Infrared Processing and Analysis Center, California Institute of Technology, 100-22, Pasadena, CA 91125, USA;\\
6: Space Telescope Science Institute, 3700 San Martin Drive, Baltimore, MD, 21218, USA;\\
7: WIYN Consortium, Inc., 950N Cherry Avenue, Tucson, AZ, 85719, USA;\\
8: Sydney Institute for Astronomy, School of Physics, University of Sydney, Camperdown, NSW 2006, Australia;\\
9: Research School of Astronomy \& Astrophysics, Mount Stromlo Observatory, Cotter Road, Weston Creek, ACT, 2611, Australia;\\
10: Institute of Astronomy, University of Cambridge, Madingley Road, Cambridge CB 0HA, UK;\\
11: Department of Astronomy, University of Arizona, 933 North Cherry Avenue, Tucson, AZ, 85721-0065, USA;\\
12: Center for the Exploration of the Origin of the Universe, Astronomy Program, Department of Physics and Astronomy, \\
Seoul National University, Seoul, Republic of Korea;\\
13: CSIRO Astronomy and Space Science, Australia Telescope National Facility, PO Box 76, Epping, NSW, 1710, Australia;\\
14: University of Michigan, 830 Dennison, 500 Church Street, Ann Arbor, MI, 48109-1042, USA;\\
15: Cerro Tololo Inter-American Observatory (CTIO), part of the National Optical Astronomy Observatory (NOAO), \\
Colina El Pino S/N, Casilla 603, La Serena, Chile;\\
16: School of Physics, University of Melbourne, Parkville, Victoria, 3010, Australia;\\
17: European Southern Observatory, Karl-Schwarzschild-Str. 2, D-85748 Garching b. M\"{u}nchen, Germany}
\date{Released 2012 Xxxxx XX}

\pagerange{\pageref{firstpage}--\pageref{lastpage}} \pubyear{2013}

\date{\today}

 \maketitle
\begin{abstract} 
H$\alpha$ observations centred on galaxies selected from the \HI\ Parkes All Sky Survey \citep[\HIPASS, ][]{Barnes2001} typically show one and sometimes two star-forming galaxies within the $\sim$15' beam of the Parkes 64-m \HI\ detections.  In our Survey of Ionization in Neutral Gas Galaxies \citep[SINGG, ][]{Meurer2006} we found fifteen cases of \HIPASS\ sources containing four or more emission line galaxies (ELGs). We name these fields Choir groups. In the most extreme case we found a field with at least nine ELGs. In this paper we present a catalogue of Choir group members in the context of the wider SINGG sample. 

The dwarf galaxies in the Choir groups would not be {individually} detectable in \HIPASS\ at the observed distances if they were isolated, but are {detected} in SINGG {narrow-band imaging} due to their membership of groups with sufficiently large { total} \HI\ mass. The ELGs in these groups are similar to the wider SINGG sample in terms of size, H$\alpha$ equivalent width, and surface brightness.

Eight of these groups have two large spiral galaxies with several dwarf galaxies and may be thought of as morphological analogues of the Local Group. However, on average our groups are not significantly \HI-deficient, suggesting that they are at an early stage of assembly, and more like the M81 group. The Choir groups are very compact at typically only 190 kpc in projected distance between the two brightest members. They are very similar to SINGG fields in terms of star formation efficiency (the ratio of star formation rate to \HI\ mass; SFE), showing an increasing trend in SFE with stellar mass.
\\
\end{abstract}

\section{Introduction}

Galaxies are arranged throughout the Universe in a hierarchy of environments ranging from clusters to groups, to isolation \citep[e.g.][]{Tully1987,Kilborn2009,Pisano2011}. Galaxies that reside within denser environments such as clusters are different to those at group densities and yet still different to those that lie in the field. The amount of star formation depends largely on the amount of gas available to fuel the process \citep{Kennicutt1989,Kennicutt1998,Bergvall2011}. Moreover, at group densities, the ratio of star-forming spiral galaxies to less prolific elliptical galaxies is lower, so morphology is important as well \citep{Wijesinghe2012}. It is not known exactly how groups transition from gas and spiral-rich to gas-poor, elliptical-rich ones like those analysed by \citet{Kilborn2009,Mulchaey1998} so the picture is incomplete. Groups of galaxies are particularly interesting because the suppression of star formation begins at group densities \citep{Lewis2002,Gomez2003}.

The selection technique for star formation studies can lead to inherent biases in the sample. Previous authors have used H$\alpha$ to select their samples \citep[e.g.][]{Gallego1995,Salzer2000}. However, H$\alpha$ followup imaging studies of optically selected galaxies are limited by the selection biases of their parent sample, typically excluding low surface brightness galaxies. The result is that these surveys are biased towards galaxies with high rates of star formation, and contain no control sample with low star formation rates.

In order to overcome that optical bias, we have selected galaxies based on their \HI\ mass measured by the \HI\ Parkes All-Sky Survey \citep[\HIPASS,][]{Barnes2001,Koribalski2004,Meyer2004}. With this sample we conducted the Survey for Ionisation of Neutral Gas Galaxies (SINGG), an \Halpha\ and R band imaging follow-up to \HIPASS.  \citet{Meurer2006} presents the SINGG sample, and gives data on 93 \HIPASS\ targets observed for SINGG. Now a total of 292 \HIPASS\ targets have been observed by SINGG with the CTIO 1.5m and 0.9m telescopes (Meurer et al. 2013, in prep).  It is these images which form the basis of this study. Fifteen fields were discovered to contain four or more H$\alpha$ sources and were dubbed Choir groups. The Choir member galaxies are different to typical field galaxies in that the larger galaxies are distorted and none are elliptical galaxies. 

In this  paper we present a catalogue of Choir group members. Section 2 outlines the sample selection and observations of SINGG. We present our catalogue of Choir group members in Section 3, along with a discussion of their properties in the context of SINGG. Section 4 concludes the paper.

We base distances on the Multipole model of \citet{Mould2000}, with $H_0=70 {\rm km s}^{-1}{\rm Mpc}^{-1}$ as in \citet{Meurer2006}. We adopt a \citet{Chabrier2003} initial mass function (IMF).

\section{Sample Selection and Data}

Our sample is drawn from the 292 \HIPASS\ targets observed for SINGG. \HI\ {measurements are all from the} \HIPASS\ \HI\ catalogue HICAT \citep{Meyer2004}, {except for two groups (}\HIPASS\ { J0443-05 and J1059-09). After noticing an anomalous} \HI\ {mass for one group, we manually remeasured the} \HI\ {mass of every Choir group. We found that the unusual } \HI\ {profiles of the Choir fields caused the automated }\HIPASS\ {parameterisation algorithm to fit poorly in these two cases. Our manually remeasured }\HI\ { masses are used in this paper for these two fields.}

The SINGG observations were mostly conducted at the Cerro Tololo Inter-American Observatory (CTIO) 1.5m telescope, whose  field of view of 14'.7 matches the $\sim15'$ beam of the Parkes radiotelescope well. Additional observations were taken at the CTIO 0.9m telescope whose field of view is 13.5'.

Emission Line Galaxies (ELGs) in SINGG were identified by eye by two of us (DH, GRM) primarily using colour composites of the SINGG data where
the red, green, and blue images of the display were assigned to the net
\Halpha\ image, the narrowband image without any continuum subtraction,
and the $R$ band image respectively.  The colour images are similar to
those shown in Figure~\ref{J0205-55}.  ELGs are
distinguished by having net line emission, and being noticeably more
extended than a point source. For unresolved emission line sources \citep[ELdots, ][]{Ryan-Weber2004,Werk2010} the distance is not clear.  They may be detached \HII\
 regions revealed by H$\alpha$ emission or background emitters of other lines (especially [O{\sc iii}] 5007) redshifted into our passband. Ancillary spectroscopy is needed to distinguish between these possibilities, and that is beyond the scope of this work; {the ELdots in the Choir fields are not discussed further in this paper.} The original data were consulted in cases
where the reality of the line emission was not clear, i.e.\ low
surface brightness or low equivalent width objects.  The images were
then measured using the standard SINGG data analysis pipeline \citep{Meurer2006}.

While most of the (\HI-rich, 15'x15') fields in SINGG contain a single ELG, there are fifteen fields that have four or more ELGs. These fields of multiple SINGGers we name Choir groups, presented in Table~\ref{groups}.

Our working assumption is that the line emission results from \Halpha\
at a velocity similar to the \HIPASS\  source, and hence that all ELGs in
a field are physically associated. {This is in the same manner as} \citet{Tully2006}{, who argued that associations of dwarf galaxies in his sample were bound.}
 For each field the narrow-band filter was chosen to most closely match the mean wavelength and wavelength range of the filter to the \HI\ velocity profile of the field. The pivot wavelengths and transmission widths are listed in Table~\ref{groups}. Typically filters with bandwidth $\sim30$\, \AA\ were used for the narrow band images of these particular SINGG fields. 
This corresponds to $\sim 3000$ \kms, much broader than the typical \HI\
line widths involved.  
Therefore spectroscopic data are needed to firmly associate all ELGs with the \HIPASS\  detection. We are in the process of confirming redshifts and these will be published in a future paper.

As this project progressed we noticed that some ELGs were missed in the
original selection of the Choir fields.  These included some small high
surface brightness galaxies as well as low surface brightness and low equivalent width (EW) detections.  We also found
cases where the morphology of a single galaxy was better described as
multiple merging or superposed galaxies.  In those cases what
distinguishes the companions as separate ELGs is a noticeable
concentration in both \Halpha\ and the $R$-band continuum.  \HII\
regions, on the other hand are distinguished by having a relatively weak
continuum above the local background and being unresolved or barely
resolved in \Halpha.

After discovering a few instances of ``new'' ELGs, one of us (GRM)
carefully examined all Choir fields, as well as SINGG fields with three
ELGs.  In total we found 13 new ELGs.  These are distinguished in
Table~\ref{choirs} by an asterisk (*).  While we think the evidence is strong that all ELGs
listed here are separate galaxies with real \Halpha\ emission, we
caution that there are some borderline cases, such as \HIPASS\ J1408-21:S6 where
the line emission has a low surface brightness and is displaced from the
parent galaxy.  While  we do believe our selection based on visual inspection is thorough, spatially varying biases and subjectivity are likely.   For example, while a strong BCD
candidate like \HIPASS\ J1051-17:S6 may be recognized even if it is projected
near a brighter companion, a small galaxy with only one or two modest
\HII\ regions, such as \HIPASS\ J0205-55a:S9 is easily noticed when isolated, but
may not be recognized as a separate galaxy if projected on or near a
bright spiral.  \Halpha\ concentrations along extended tidal arms, such
as \HIPASS\ J1250-20:S5,S6 are especially ambiguous.  It is not clear whether they are separate tidal
dwarf galaxies \citep[e.g., ][]{Bournaud2007} or just transitional \HII\
regions.

The new ELGs in the sample were not measured using the SINGG
measurement pipeline, since it was not operational when the measurements were required.  Instead
basic mesaurements of position and fluxes were measured using {\sc
  imexam} in {\sc iraf}\footnote{{\sc iraf} is distributed by the National Optical Astronomy Observatories, which are operated by the Association of Universities for Research in Astronomy, Inc., under cooperative agreement with the National Science Foundation \citep{Tody1993}.}.

In summary, the following criteria must be met to satisfy our Choir group definition:
\begin{enumerate}
\item \HI\ detection in \HIPASS;
\item four or more ELGs in a single field of view of $\sim$15';
\item where an ELG is defined by net H$\alpha$ emission in an extended source. 
\end{enumerate}

{We point out that the above is the \emph{minimum} to define a Choir group. The definition has the following caveats:}
\begin{enumerate}
\item{{Choir groups can be larger than 15', with members outside of the field of view;}}
\item{{Choir groups can therefore belong to much larger structures, e.g. }\HIPASS\ { J0400-52, which is in Abell 3193};}
\item{{Choir groups require spectroscopic followup to confirm assumed physical association.}}
\end{enumerate}
 
  {These caveats are discussed more fully in Section 3.}
 
We present the Choir groups in Table~\ref{groups}, and key properties of the individual Choir group members in Table~\ref{choirs}. These data are preliminary results on all the galaxies observed with the CTIO 1.5m and 0.9m telescopes for SINGG. Full results are in preparation and will be presented elsewhere (Meurer et al., in prep).

\begin{figure*}
\centerline{
\includegraphics[width=1\linewidth]{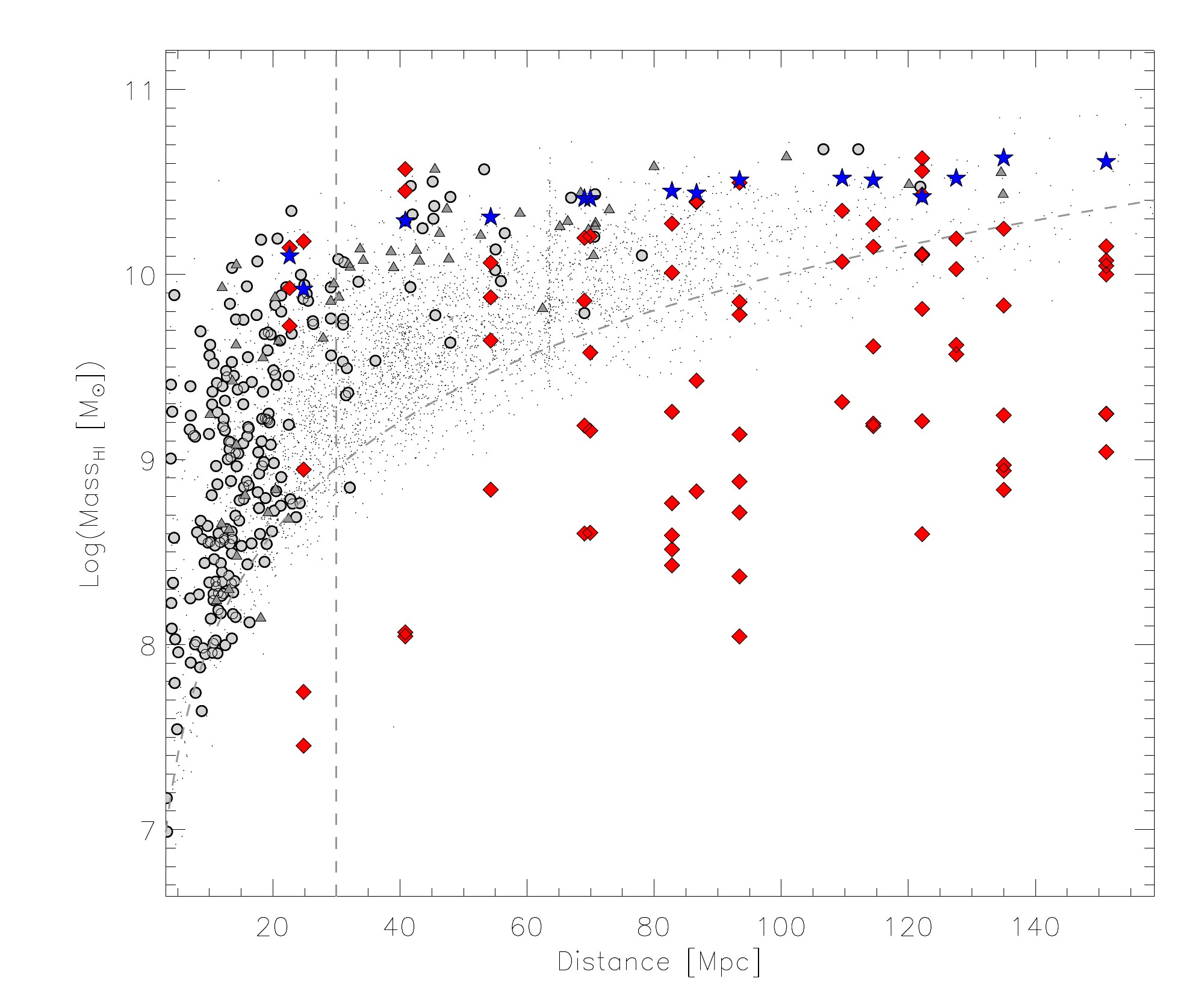}
}
\caption{Total \HI\ mass vs. distance for SINGG detections compared to \HIPASS. The blue stars are Choir groups, mid grey triangles denote doubles and triples, and mid grey filled circles are single galaxies in SINGG. Black points indicate \HIPASS\ detections not in SINGG. The vertical, dashed line at 30 Mpc represents the field of view limit for detecting an average Choir group. See text for explanation. The curved, dashed line represents the 3$\sigma$ detection limit in \HIPASS\ as described in \citet{Zwaan2004} and \citet{Meurer2006}, from a fake source analysis and integrating over all line widths from 20 to 650 km s$^{-1}$. Choirs are at the high M$_{HI}$ and large distance end of the distribution. 
Estimated \HI\ mass for Choir member galaxies is shown as red diamonds. See text for calculation. Above the nominal group detection limit of 30 Mpc, only the brightest members of each group are detectable at the 3$\sigma$ limit if isolated: the Choir dwarfs are only detected due to their inclusion in an \HI-rich group. 
\label{distanceVlogmhirain}}
\end{figure*}

\section{Discussion}

All the galaxies in SINGG have (by design) \HI\  and all are detected in H$\alpha$, indicating that \HI-rich, non-star-forming galaxies are rare \citep{Meurer2006}. Fields observed for SINGG usually contain single ELGs, with some doubles and triples, and more rarely four or more galaxies in a single pointing (our Choir groups). We use the entire SINGG dataset as our control sample against which we compare the Choirs galaxies. In this section, we discuss selection biases, analyse the Choir member galaxies in terms of size, equivalent width, luminosity and surface brightness, and then focus on the Choir groups' morphology, size, star formation rate and efficiency, and \HI\ deficiency.

\subsection{Selection biases} 
Although SINGG overcomes biases that are prevalent in optically-selected surveys, some selection effects are still present. The two major selection effects are 1. a selection of more massive sources, and 2. a bias towards more distant groups.

First, the SINGG sample is selected from \HIPASS\ so that the nearest sources at each \HI\ mass are preferentially chosen; combined with the \HIPASS\ \HI\ detection limit this means that distant, isolated, low-mass \HI\ sources are not selected (see Figure~\ref{distanceVlogmhirain}). This is therefore also a selection effect for Choir groups. The detection limits are discussed in detail by \citet{Zwaan2004}. 
At higher redshift (distance $\gtrsim$30 Mpc) only the most massive \HI\ sources are detected by \HIPASS. These sources are so rare that we can not find many of these except by looking at these distances. Hence most of the high mass $M_{\rm HI} > 10^{10} M_\odot$ sources selected for SINGG have $D > 30$ Mpc.
SINGG can detect galaxies optically to fairly low stellar masses out to the full $\sim$150 Mpc distance limit of \HIPASS. While the \HI\ mass detection limit precludes us from detecting isolated dwarf galaxies at distances greater than about 30 Mpc, we can detect them at these distances when they are part of a more massive \HI\ system. We illustrate this in Figure~\ref{distanceVlogmhirain}. Choir groups (blue stars), SINGG doubles and triples (grey triangles) and SINGG singles (light grey circles) all show increasing \HI\ mass with distance. 

\begin{figure}
\centerline{
\includegraphics[width=1\linewidth]{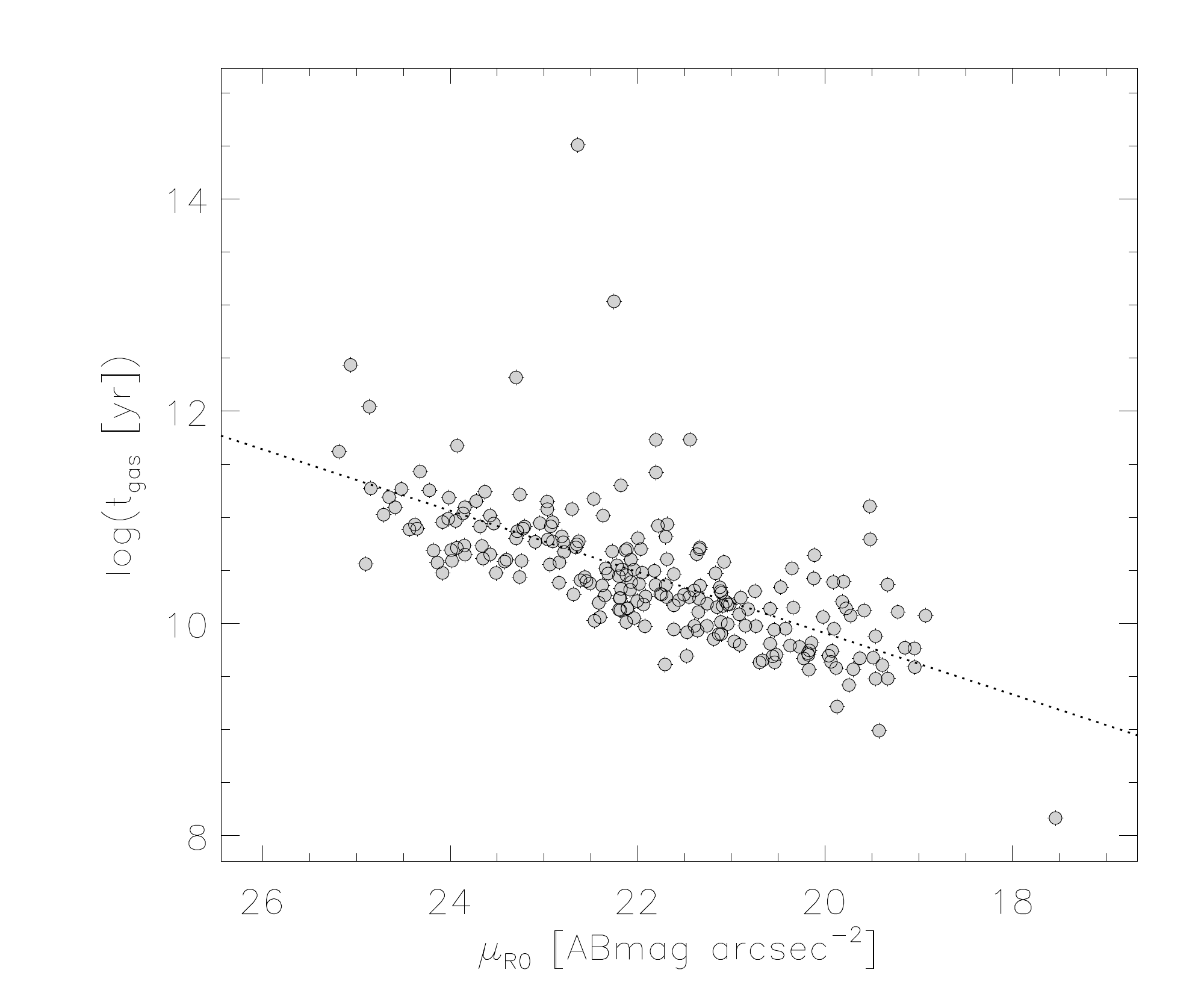}
}
\caption{Parameterisation of gas cycling time as a function of $R$-band effective surface brightness for single galaxies in SINGG. 
\label{tgas}}
\end{figure}

In order to show the likely contribution of the individual galaxies within the Choirs to the system \HI\ mass, we bring some basic correlations seen within SINGG to bear. Following \citet{Meurer2006} we define the gas cycling time $t_{gas} {\rm [yr]} = 2.3M_{HI}/SFR$, where $M_{HI}$ is the \HI\ mass and the factor of 2.3 is a correction for molecular hydrogen and helium content.
We then adopt the \citet{Meurer2009} conversion of star formation rate SFR [M$_\odot$ yr$^{-1}] = L_{H\alpha}/(1.5\times1.04\times 10^{41})$, where $L_{H\alpha}$ is the $H\alpha$ luminosity in ergs/s. The factor of 1.5 converts  the \citet{Salpeter1955} IMF measurements of SINGG to a \citet{Chabrier2003} profile \citep[][]{Brinchmann2004,Meurer2009}. In Figure~\ref{tgas} we plot $t_{gas}$ as a function of $R$-band effective surface brightness $\mu_e$, with the best fit 
\begin{align}
\log(t_{gas}) = (4.14\pm0.48) + (0.29\pm0.02)\mu_e.
\end{align}
This allows us to estimate $M_{HI}$ from $L_{H\alpha}$ and $\mu_e$ as follows:
 \begin{align}
 \log(M_{HI}) = \log(L_{H\alpha}) + (0.29\pm0.02)\mu_e - (37.42\pm0.48).
 \end{align}
The significance of this relation will be discussed in the context of SINGG in a future paper.

We use this relation to predict the \HI\ mass of individual Choir member galaxies, shown as red diamonds in Figure~\ref{distanceVlogmhirain}. If the galaxies were isolated, only the brightest galaxies in each Choir group could be detected in \HI\ by Parkes. The smaller members of the Choir groups could not be detected, and are only included in SINGG due to their inclusion in an \HI-rich group. {The groups at 40 and 120Mpc, }\HIPASS\ {J1403-06 and J1059-09, each have a total \emph{observed}} \HI\ {mass less than the \emph{predicted} }\HI\ {mass of their two-three brightest members. This means that these groups are both deficient in} \HI\ {compared with the amount expected based on the H$\alpha$ luminosity and $R-$band surface brightness of their group members. See Section~}\ref{HIDef} {for a discussion of }\HI\ {deficiency.}

The second selection effect is a bias towards more distant groups; there are fewer Choir groups and fewer members per group detected at small distances. This is because the large angular size of nearby groups is more likely to exceed our 15' field of view. A single pointing will then contain fewer than all of the members in a group, leading to underrepresentation of the number of galaxies identified as group members. (We note previously detected giants that are likely to be asssociated with our Choir groups in Appendix A.) Also, if a pointing contains less than four objects (the threshold for defining a Choir), a group will not be detected, leading to underrepresentation of number of groups at small distances. 

For a Choir group to be detected, it must have at least four ELGs within the field of view. We characterise group size by measuring the projected distance between the two most luminous galaxies in each group. (See Section~\ref{compactness}.) Our mean Choir group size is 190 kpc, which will fit inside a single pointing as near as $\sim$ 30 Mpc. Therefore, we do not expect to see any groups of this average size nearer than 30 Mpc (represented by the vertical dashed line in Figure~\ref{distanceVlogmhirain}). This corresponds closely with our observations; although there are two groups below this cut, one is very compact (\HIPASS\ J1159-19) and the other barely makes the Choir definition with one member nearly outside of our field of view (\HIPASS\ J2318-42a).

It is important to note that many of the nearby SINGG galaxies are likely to be in groups where only three or fewer galaxies fit within the SINGG field of view. We estimate the fraction of SINGG that is in groups similar to the Choirs by measuring the proportion of Choir groups compared with all SINGG detections at distances greater than 30 Mpc. In this manner we calculate that 20\% of SINGG detections are in fact in galaxy groups.  Considering that Choir groups are still likely to be under-represented at the near end of this distance range, the true fraction may be significantly higher. The proportion of groups increases with distance. According to \citet{Tully1987}, around 50\% of galaxies are expected to be in groups of 4 or more members. 

\begin{figure}
\centerline{
\includegraphics[width=1\linewidth]{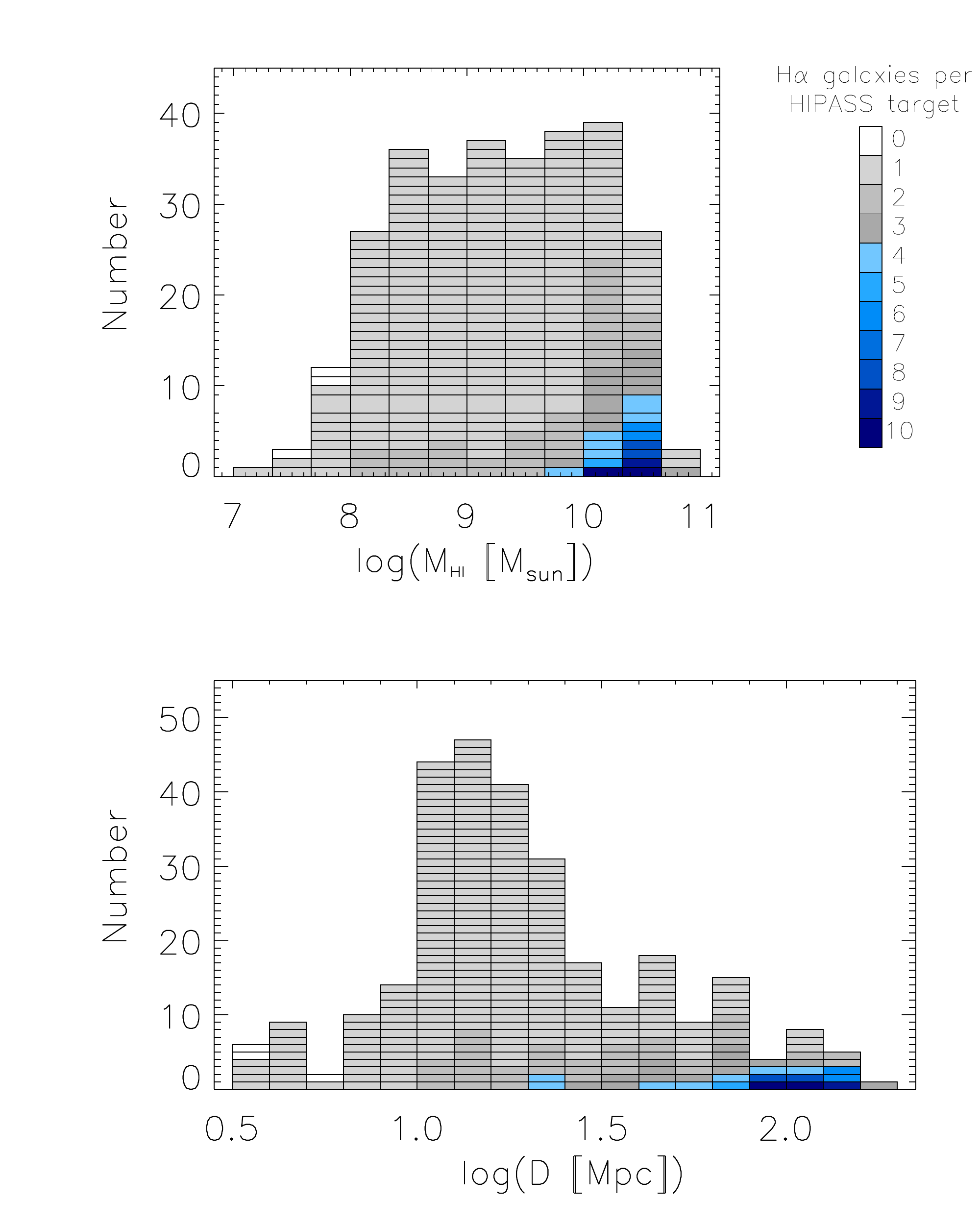}
}
\caption{HI mass histogram and distance histogram of SINGG detections. Blue colours correspond to Choir groups; darker colours correspond to more members (see key). These histograms are nested, so that the entire area covers the whole SINGG sample. Choirs are at the high M$_{HI}$ and large distance end of the distribution.
\label{HIhist}}
\end{figure}

These two selection effects mean that Choirs are among the most distant and \HI-massive of the \HIPASS\ sources. Figure~\ref{HIhist} illustrates the distribution of groups in both \HI\ mass and distance. While the  SINGG control sample has (by design) a relatively flat distribution between $8 < log(M_{HI}) < 10.6$, the number of Choir groups peaks at the high-mass end of this range. These differences must be taken into account when comparing Choirs with the control SINGG sample for distance-dependent and mass-dependent quantities.

In the following subsections we continue this discussion with an analysis of the properties of the Choir member galaxies.

\subsection{Size and Equivalent Width}

\begin{figure}
\centerline{
\includegraphics[width=1\linewidth]{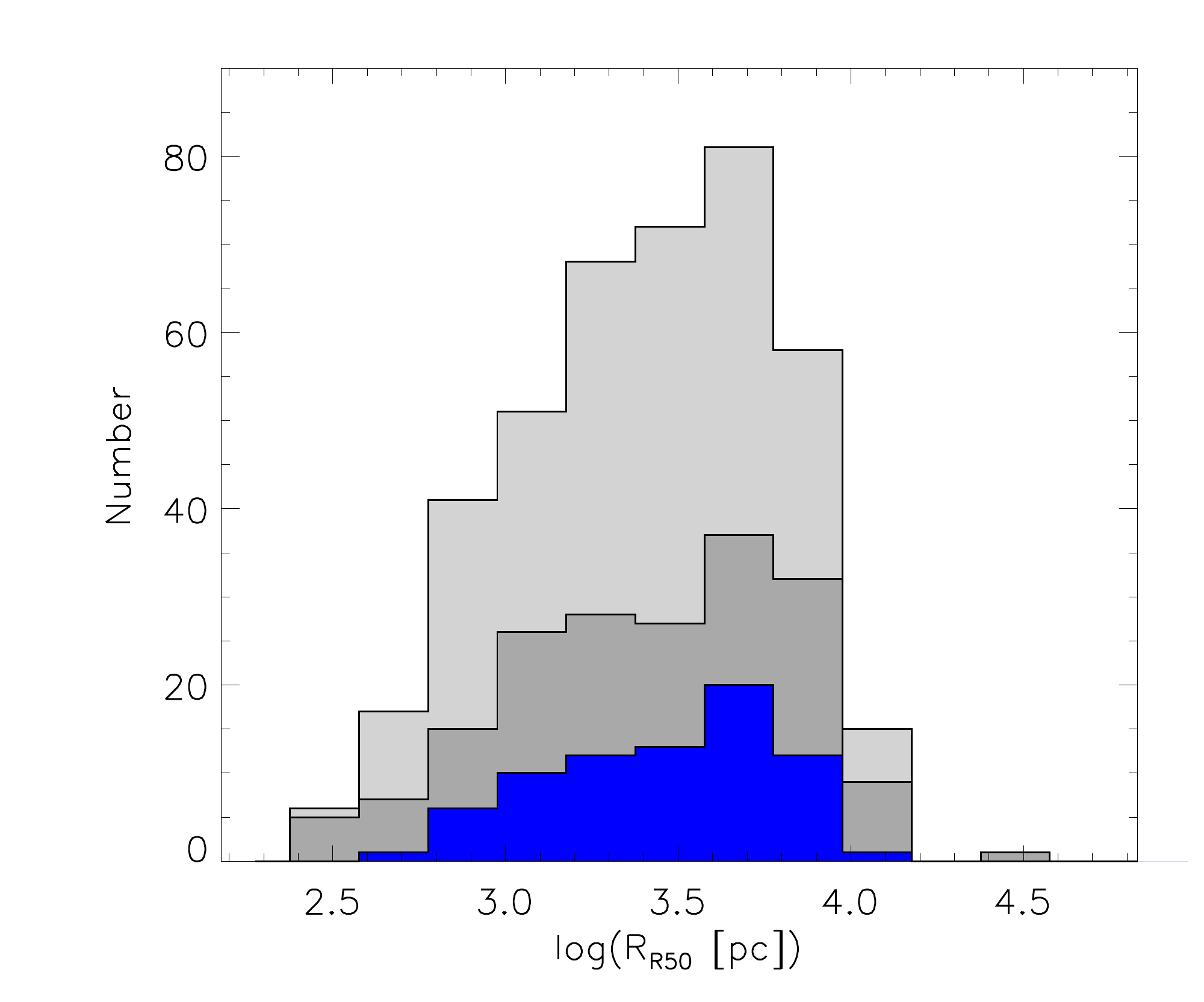}
}
\caption{Histogram of $R$-band half-light radius of ELGs in SINGG. Blue, mid grey and light grey denote Choir member galaxies, SINGG doubles and triples, SINGG single galaxies respectively. Choir members are not significantly different from the control SINGG sample ($p=0.35$). This is similar for H$\alpha$ half-light radius, $R$-band radius enclosing 90\% of flux, H$\alpha$ radius enclosing 90\% of flux. The same is seen when R $>$ -21 to compare only dwarf galaxies. 
\label{re_r_t_all}}
\end{figure}

The histogram of $R$-band effective (half-light) radius, $r_e(R)$ for  Choir member galaxies, in comparison to other single and multiple SINGG galaxies, is shown in Figure~\ref{re_r_t_all}. A Kolmogorov-Smirnov (KS) test shows that Choir members are not significantly different from the single detections in SINGG with a fractional probability that they were drawn from the same parent sample of $p = 0.35$. A similar result occurs for H$\alpha$ effective radius, radius enclosing 90\% of $R$-band flux and  radius enclosing 90\% of H$\alpha$ flux. Figures demonstrating this are not shown for the sake of brevity. Applying a magnitude cut at $M_R > -21$ to exclude the most luminous galaxies does not alter the result; lower luminosity Choir galaxies are also not significantly different from their SINGG counterparts.

\begin{figure}
\centerline{
\includegraphics[width=1\linewidth]{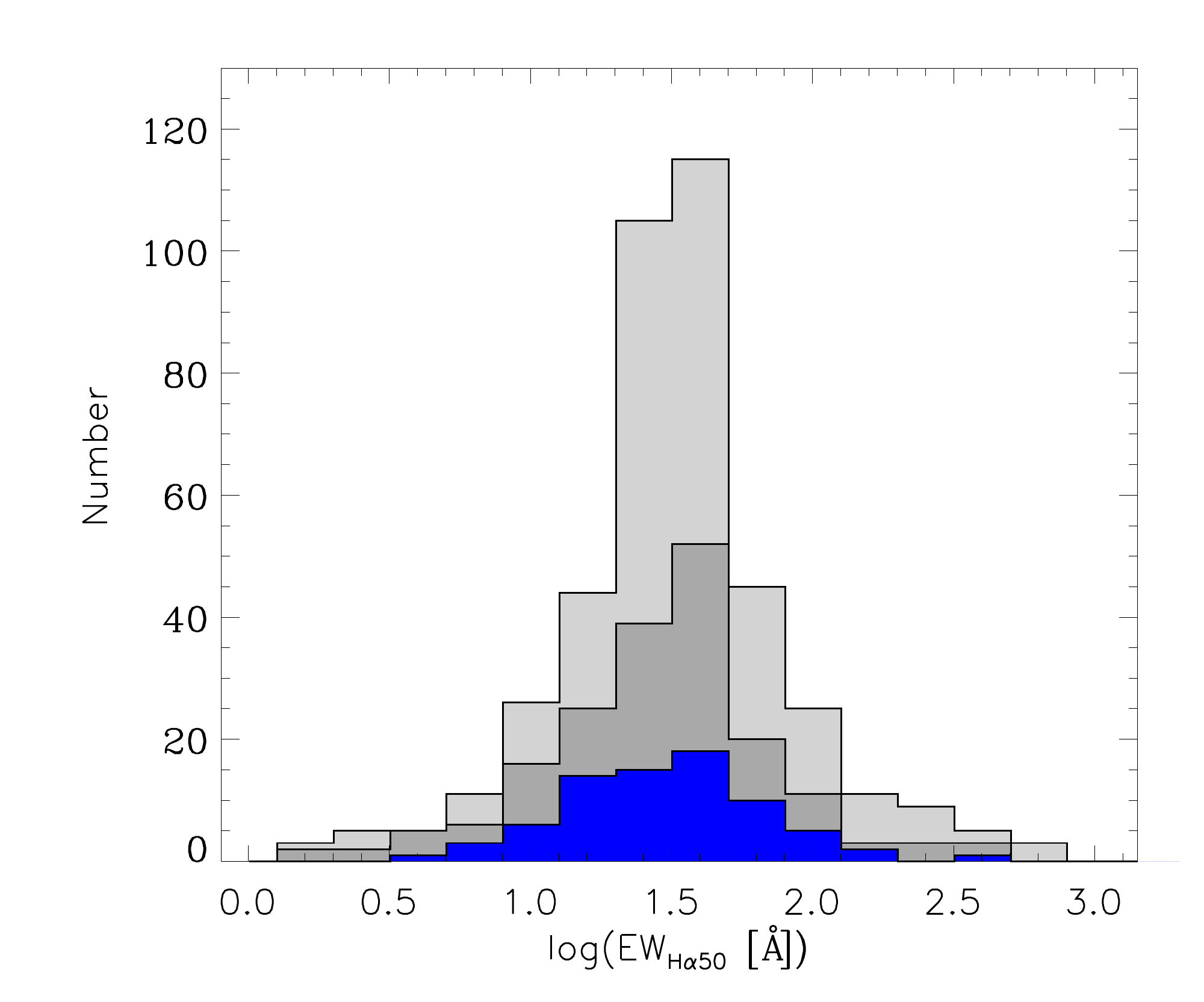}
}
\caption{H$\alpha$ EW calculated within the effective radius and corrected for dust. Blue, mid grey and light grey denote Choir member galaxies, SINGG doubles and triples, SINGG single galaxies respectively. Choir members do not have high EW for their size ($p=0.54$). The same is seen when R $>$ -21 to compare only dwarf galaxies).
\label{ew50_0_t}}
\end{figure}

Figure~\ref{ew50_0_t} is a histogram of H$\alpha$ equivalent width (measured within the H$\alpha$ effective radius and corrected for dust absorption). Choir members do not have high EWs when compared with the SINGG control sample ($p=0.54$). The same result is seen for lower luminosity galaxies ($M_R > -21$).

Naively, one might expect the distance-dependent detection limit in \HI\ mass, together with the fact that Choirs are at further distances, to cause a dependence of radius and EW on distance as well. However, as discussed above, Choir dwarfs are included in the SINGG field of view only because of their proximity to \HI-detectable giants. We have used the Choir groups to identify star-forming dwarfs at such large distances that they are not detectable in \HIPASS, but their optical properties are the same as nearby star-forming dwarfs detected in \HIPASS.

\subsection{Luminosity and Surface brightness}

\begin{figure}
\centerline{
\includegraphics[width=1\linewidth]{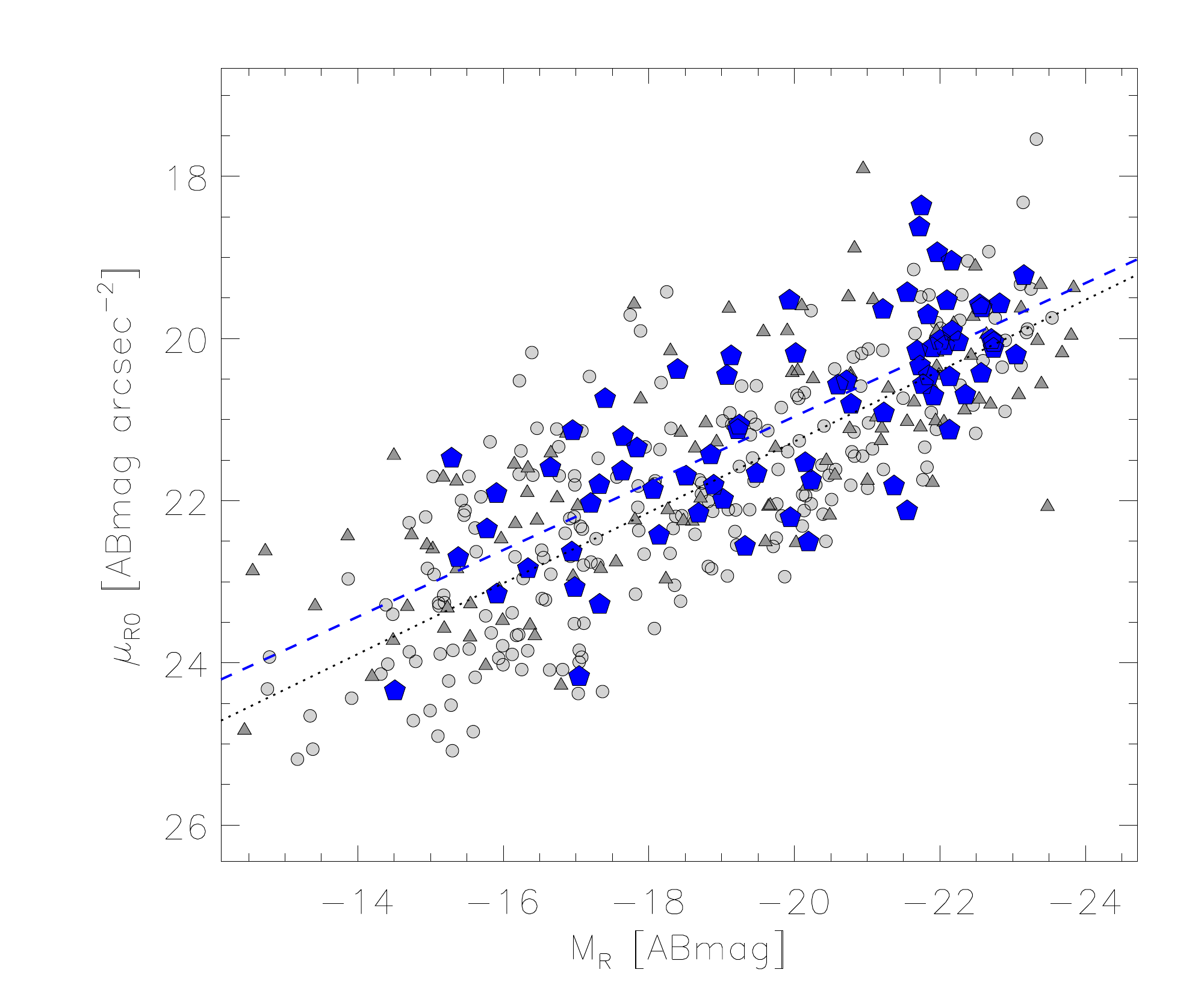}
}
\caption{Surface brightness in $R$-band as a function of absolute magnitude. The blue pentagons are Choir member galaxies, mid grey triangles denote doubles and triples, and mid grey filled circles are single galaxies in SINGG. The blue, dashed line is a linear fit to Choir members and the dotted line is for single galaxies in SINGG. The small offset is not significant ($p=0.06$).
\label{mabs_r0_tVmu_e_r0_t}}
\end{figure}

\begin{figure}
\centerline{
\includegraphics[width=1\linewidth]{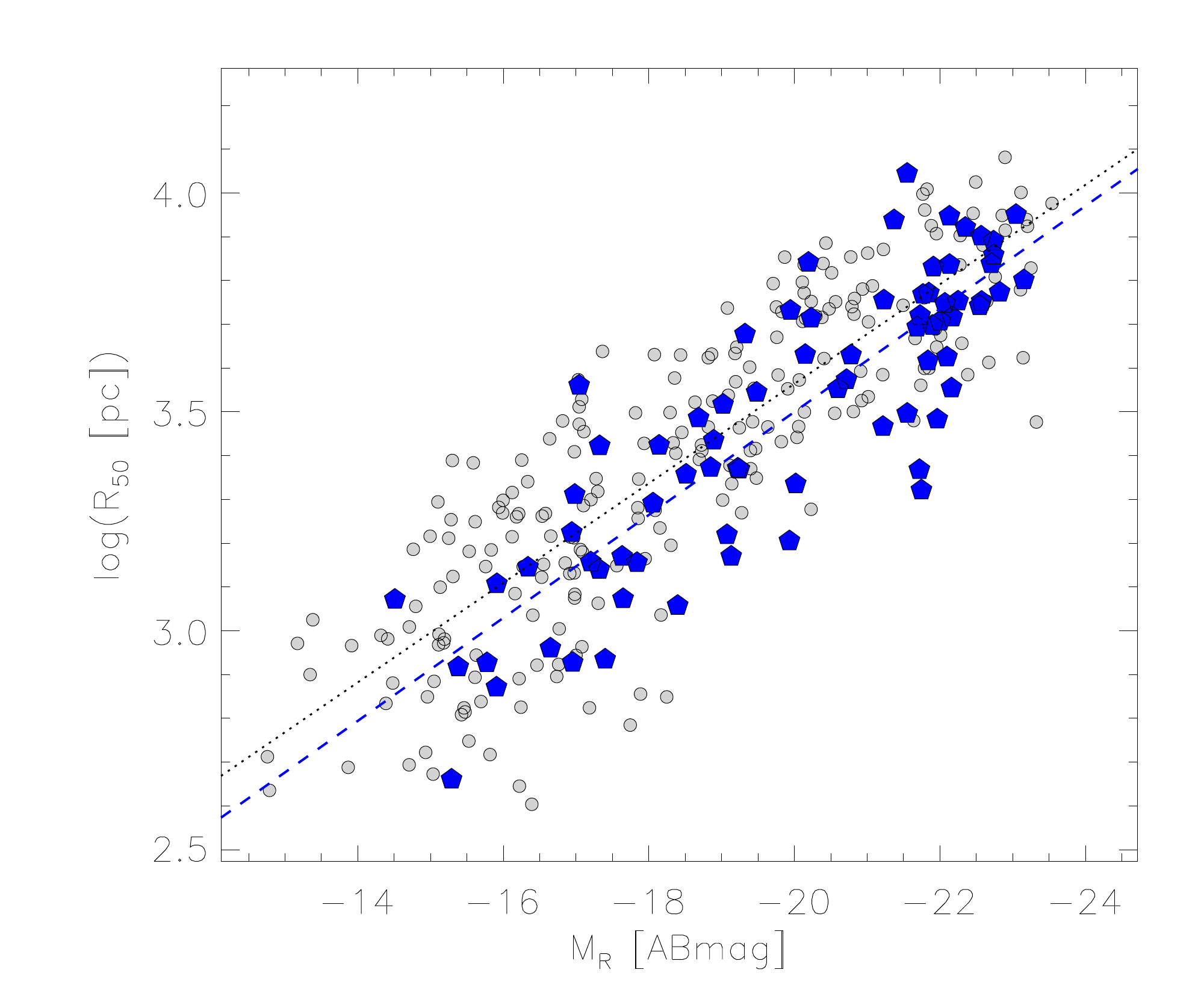}
}
\caption{$R$-band half-light radius as a function of absolute magnitude. The blue pentagons are Choir member galaxies, mid grey triangles denote doubles and triples, and mid grey filled circles are single galaxies in SINGG. The blue, dashed line is a linear fit to Choir members and the dotted line is for single galaxies in SINGG. The small offset is not significant ($p=0.27$).
\label{mabs_r0_tVre_r_t}}
\end{figure}

In Figures~\ref{mabs_r0_tVmu_e_r0_t} and~\ref{mabs_r0_tVre_r_t}  we plot luminosity-surface brightness and luminosity-radius correlations. Choir galaxies have on average 0.5 dex higher surface brightness and 0.05 dex smaller radius for their luminosities than the control sample. We perform a KS test on the distribution of $\{y - (a + bx)\}$, where $y$ is the surface brightness or radius and $x$ is the $R$-band magnitude of the Choir galaxies, and $a$ and $b$ are parameters from the fit to single galaxies in SINGG. We find that the offsets are not significant, with p-values of 0.06 and 0.27 respectively.

\subsection{Group morphology}

\begin{figure}
\centerline{
\includegraphics[width=1\linewidth]{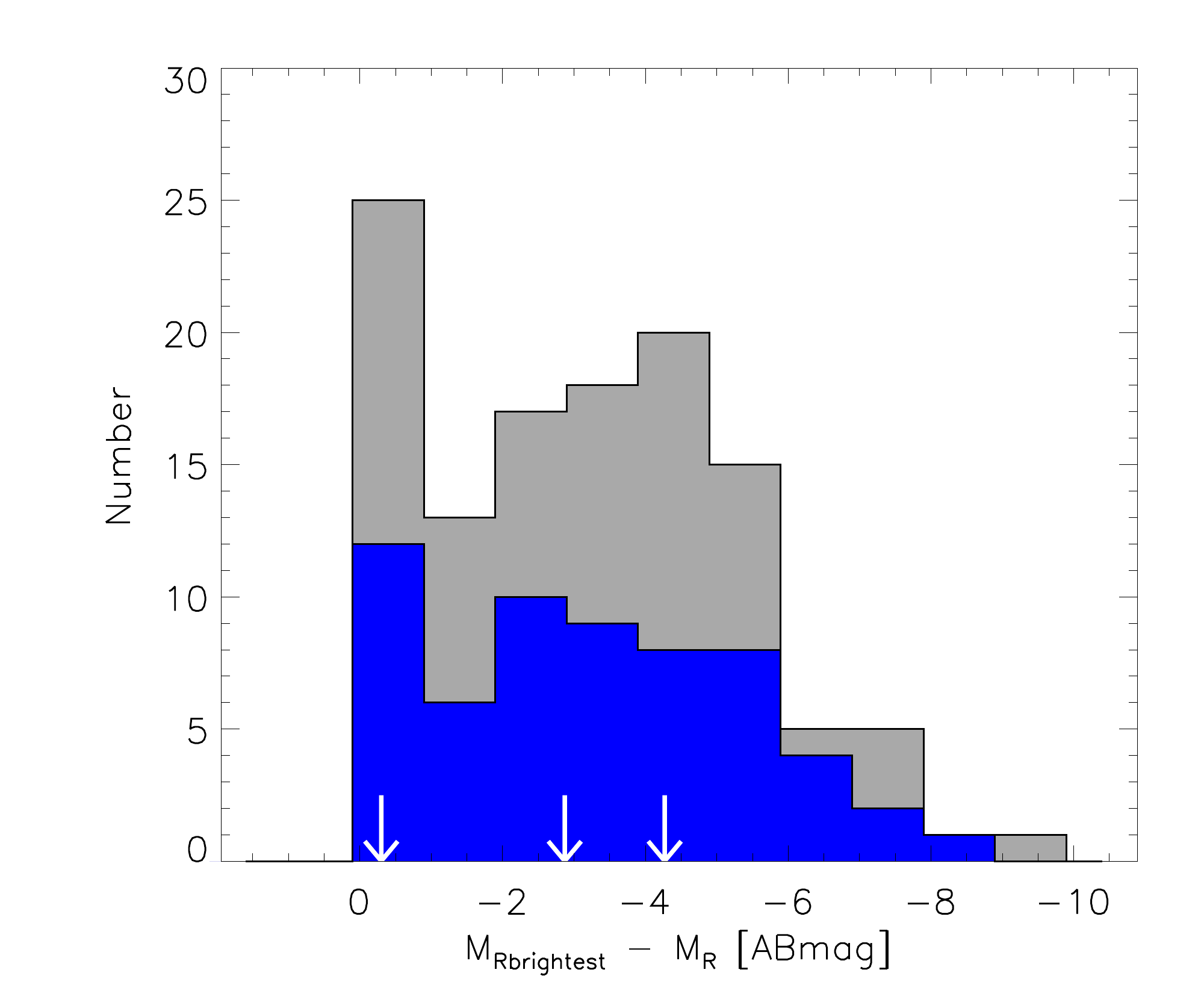}
}
\caption{Distribution of relative luminosities of group galaxies compared to the most luminous in each group. The blue area denotes Choir member galaxies, mid grey denotes doubles and triples in SINGG. Single galaxies in SINGG are not shown, as there are no fainter companions in these fields. The first peak indicates when there are two large galaxies in a group and the second broader peak shows the dwarf members. The white arrows denote the position of Local Group members relative to M31 (left to right: MW, LMC, SMC). Qualitatively, our groups have a similar distribution of relative $R$-band magnitude to our Local Group.
\label{mabs_r0_tREL}}
\end{figure}

The Choir groups by definition have four or more H$\alpha$-emitting galaxies, without further restriction on morphology or relative size. An interesting subset (eight out of fifteen groups) is groups that are comprised of two large spirals and two to eight smaller galaxies. We illustrate this in Figure~\ref{mabs_r0_tREL} where we show $R$-band absolute magnitude of Choir  members relative to the brightest member in each group. The peak at -0.25 mag represents the second-largest giant, and the extended tail peaking at -2.25 mag represents dwarf companions. We note that $M_r$ magnitudes for the Milky Way (MW), M31, Large Magellanic Cloud (LMC) and Small Magellanic Cloud (SMC) are  -21.17, -21.47, -18.60 and -17.20 respectively \citep{Robotham2012}, so that the Local Group will appear on this plot at the white arrows.
 In terms of luminosity the Choir groups therefore appear to be possible Local Group (LG) analogues, as discussed by \citet{Pisano2011,Robotham2012}. Our selection method seems to be good at finding LG analogues (at least in terms of {\emph{magnitude}} and {\emph{morphology}}), with an approximate strike rate of 50\%. We suggest that perhaps these types of groups are more common than previously thought, but usually the dwarf galaxies fall below the relevant detection limit so the group appears as a pair of bright spirals. In SINGG however, star-forming dwarf galaxies are readily apparent in the H$\alpha$ imaging.

In Appendix A we point out some morphological features of each Choir group {and search larger photographic survey images to check for possible group members outside of our imaging.} Interestingly, there are no {\emph{bright}} ellipticals in the {SINGG imaging, and the few nearby giant ellipticals do not appear to be associated with the }\HIPASS\ {detections. This is in contrast with} optically- and X-ray-selected groups where the elliptical fraction is 0.4 to 0.5 \citep{Mulchaey2000}. The discrepancy is probably a consequence of the \HI\ selection in \HIPASS\ being biased towards younger, \HI-rich groups with fewer ellipticals.

\subsection{Group compactness\label{compactness}} 

In this section we compare the size of our Choir groups to Hickson Compact Groups \citep[HCGs, ][]{Hickson1989} and groups in the \citet{Garcia1993} catalogue. These three catalogues all contain groups of four or more members, but have different limiting magnitudes and distance ranges{, and different group-finding algorithms}.

Ideally, galaxy group size is measured by {the virial radius defined as the radius enclosing a luminosity} brighter than a specified magnitude \citep[e.g., ][]{Tully1987,Garcia1993,Garcia1995}. This measurement requires radial velocity data, which do not yet exist for most of our Choir group members. It also assumes a relaxed group with a Gaussian distribution of velocities, but our Choir groups are not relaxed and do not have a sufficient number of members to display a Gaussian distribution. We are limited by having only a few members, particularly in the majority of cases where there are only two bright spirals and a number of faint dwarfs. While it may appear possible to use the projected distance between two closest neighbours in the group to compare our groups to other samples, this statistic should only be used to compare catalogues that have consistent limiting magnitudes, which is not the case for Choirs, HCGs and Garcia groups. We therefore use the projected distance between the two \emph{most luminous} galaxies in each group as a measurement of `group compactness'. This parameter {is not} as physical as previously-mentioned measurements of group size, but {simply} allows us to put our groups in context with existing catalogues given the available data. {We emphasise that our comparison is not strict, because the catalogues are based on different algorithms.}

\begin{figure}
\centerline{
\includegraphics[width=1\linewidth]{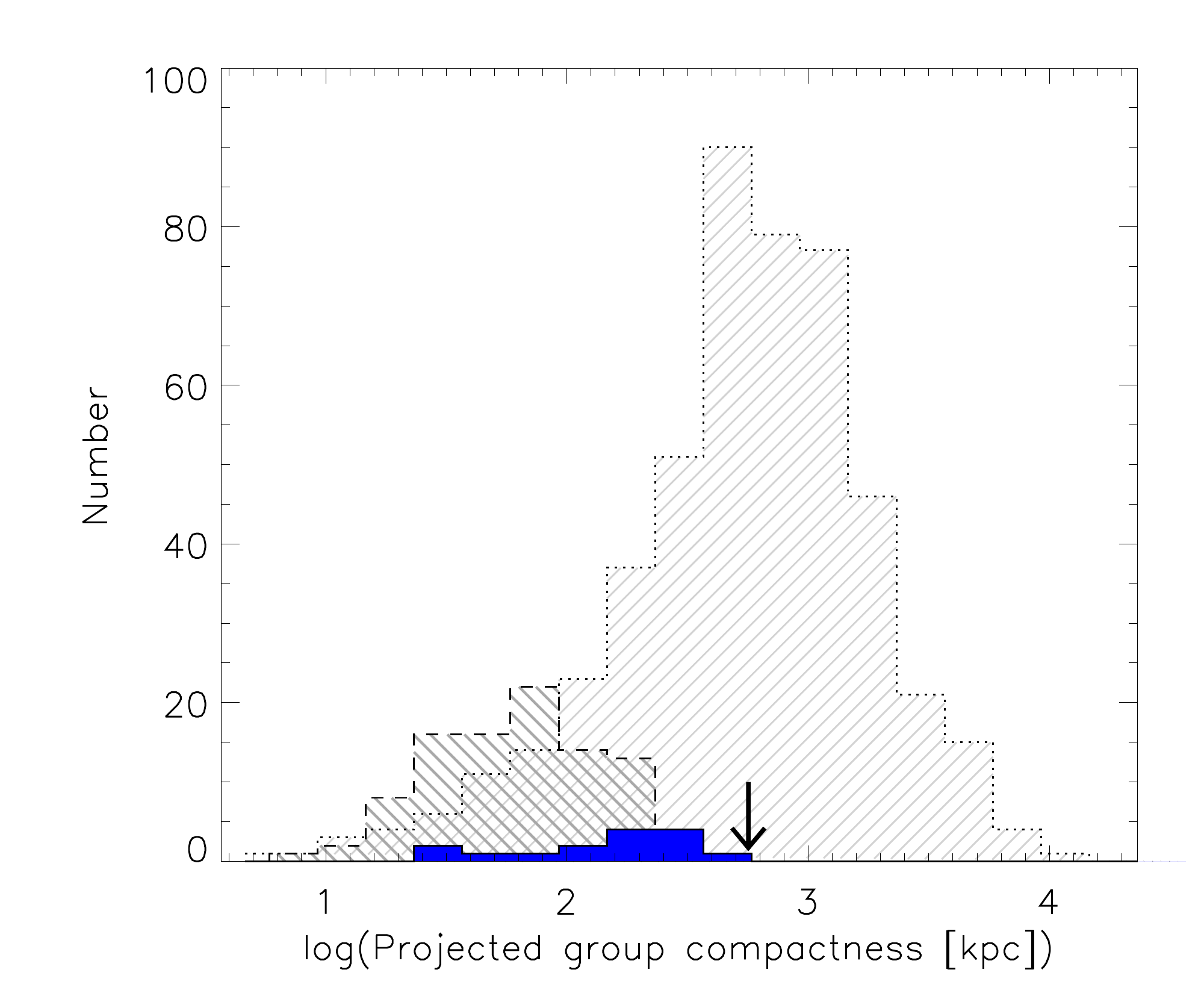}
}
\caption{Choir group compactness, estimated by measuring separation between two brightest galaxies in a group.  The solid, blue histogram is our Choir groups; light grey SW-NE cross-hatching with dotted outline is Garcia groups; medium grey NW-SE cross-hatching with dashed outline is Hickson groups. Our groups are more compact than Garcia groups, but not as compact as Hickson groups. The black arrow indicates the compactness of our Local Group, which is more than 3$\sigma$ from the mean Choir group compactness.
\label{group_hist}}
\end{figure}

For each of the three catalogues we calculate the compactness parameter and show histograms for the different catalogues in Figure~\ref{group_hist}. Mean group compactnesses for Choir groups, Hickson groups and Garcia groups are 190$\pm$31, 87$\pm$8 and 961$\pm$52 kpc respectively. The distributions are significantly different; a KS test yields $p < 0.001$ that Choir groups and Garcia groups belong to the same population, and $p <0.001$ that Choir groups and Hickson groups belong to the same population. Of course, Choir group sizes are limited by the field of view of the CTIO images, causing our distribution to be skewed in favour of smaller groups;  at our mean distance of 87 Mpc the maximum size of our groups is only 380 kpc.

For the Local Group this group compactness statistic is 800 kpc in 3D space. Using the typical $\sqrt{2}$ conversion factor, this corresponds to 565 kpc in 2D space. This is just over 3$\sigma$ larger than our mean Choir group compactness. In terms of physical separation then, we note that Choirs appear to be a compressed version of the Local Group, and may represent a later stage of evolution of a system like M31 and the MW with their retinue of dwarfs. 

A more sophisticated analysis that includes radial velocity measurements for a stricter definition has recently been conducted for the Galaxy And Mass Assembly sample \citep[GAMA, ][]{Robotham2012}, with the result that LG analogues are rare in that sample. We plan to conduct a similar analysis of the frequency of LG analogues in SINGG.

\subsection{Star formation}

\begin{figure}
\centerline{
\includegraphics[width=1\linewidth]{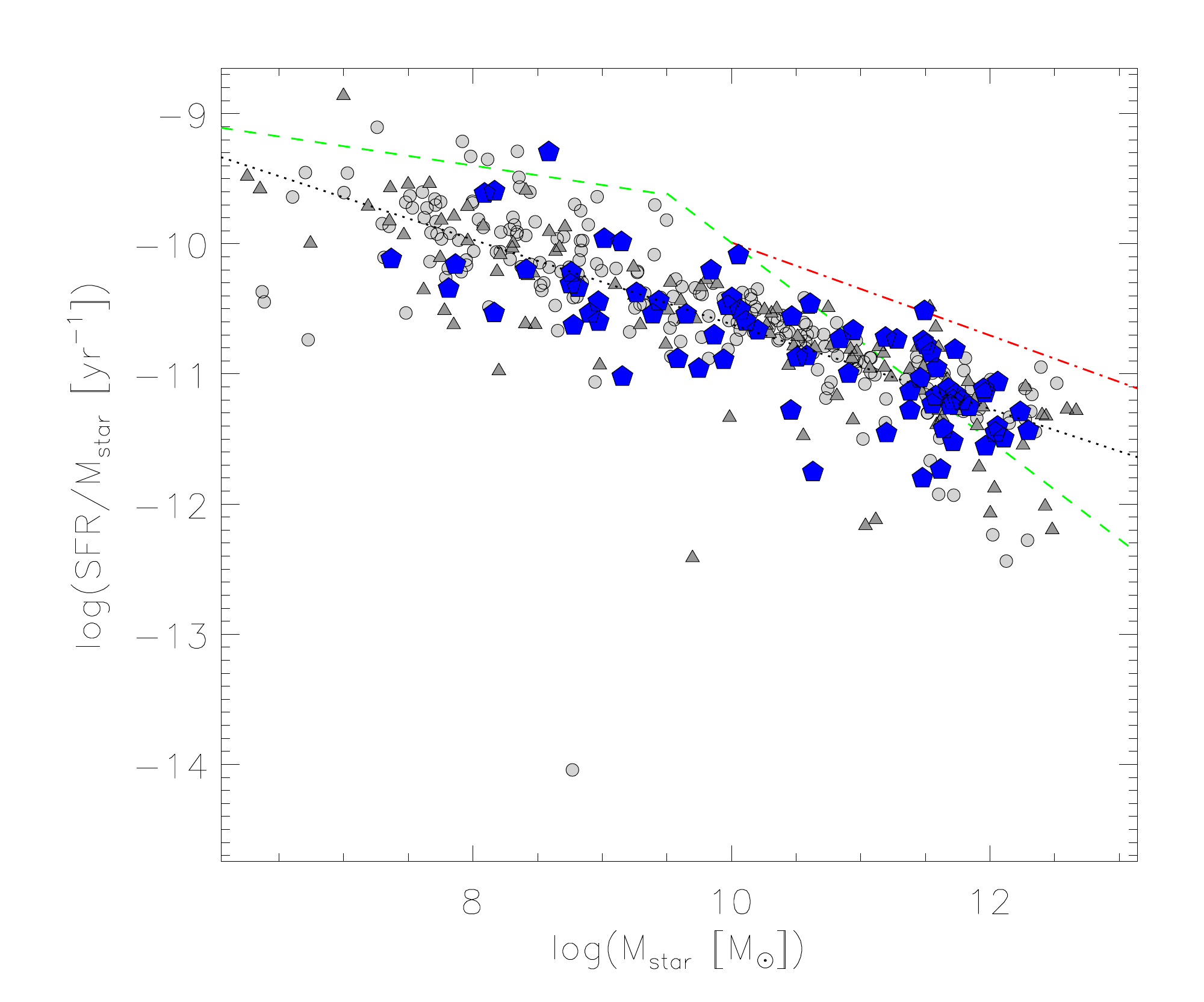}
}
\caption{Specific star formation rate as a function of stellar mass for individual galaxies in SINGG. The blue pentagons are Choir member galaxies, mid grey triangles denote doubles and triples, and mid grey filled circles are single galaxies in SINGG. The black, dotted line is the best fit to single galaxies in SINGG. The red, dot-dashed line is the best fit to high-sSFR galaxies in \citet{Schiminovich2010}. The green, dashed line is the \citet{Huang2012} relation. Choir galaxies lie on the relation defined by the control SINGG sample ($p=0.37$). The SINGG sample exhibits a lower specific star formation rate than the Schiminovich sample across all stellar masses ($p<0.001$). The high stellar mass Huang relation is better matched to the SINGG sample but displays a much steeper slope.
\label{MstarVSFRonMstar}}
\end{figure}

\begin{figure}
\centerline{
\includegraphics[width=1\linewidth]{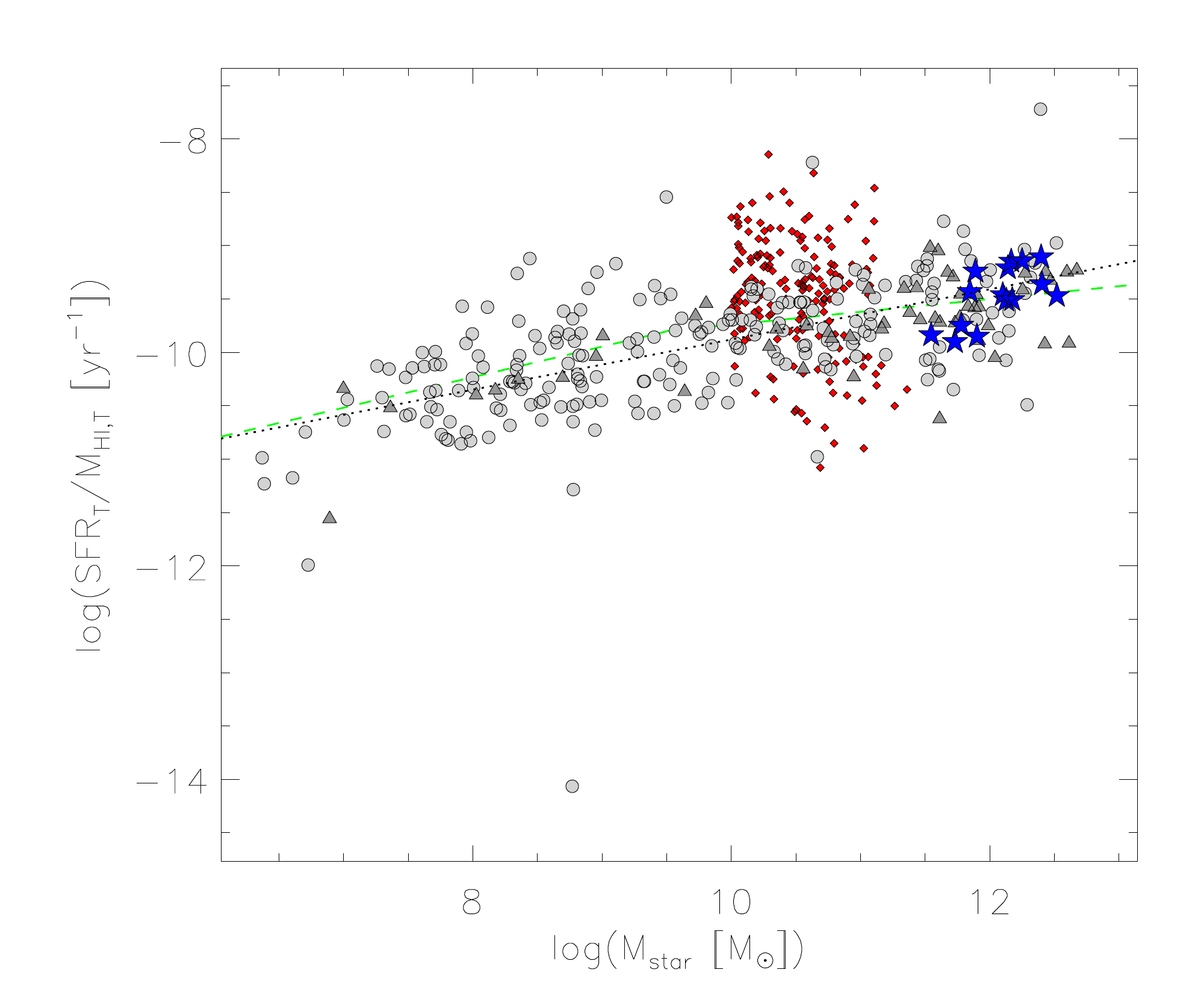}
}
\caption{Total star formation efficiency as a function of total stellar mass for groups in SINGG. The blue stars are Choir groups, mid grey triangles denote doubles and triples, and mid grey filled circles are single galaxies in SINGG. The black, dotted line is for single galaxies in SINGG. The small, red diamonds are the high-sSFR galaxies in \citet{Schiminovich2010}. The green, dashed line is the ridge line of the \citet{Huang2012} sample. Choir groups lie on the relation defined by the control SINGG sample ($p=0.14$). The SINGG sample has a lower star formation efficiency than the high-sSFR Schiminovich sample within the corresponding stellar mass range ($p<0.001$).
\label{MstarVSFRonlogmhi}}
\end{figure}

In Figures~\ref{MstarVSFRonMstar} and~\ref{MstarVSFRonlogmhi} we plot specific star formation rate (sSFR) and total (group) star formation efficiency (SFE$_T$) as a function of stellar mass $M_*$, where sSFR = SFR/M$_*$ and SFE$_T$ = SFR$_T$/M$_{HI,T}$. The subscript $_T$ denotes total quantities for each group. Stellar masses are estimated using the \citet{Bell2003} conversion $\log(M_*/L_g) = -2.61 + 0.298\log(M_*h^2/M_\odot)) $, with $M_{R\odot} = 4.61$, $M_{g\odot} = 5.45$ and ($g-r$) = 0.5 mag for late-type galaxies \citep{Blanton2003}. This gives $\log(M_*) = -3.66 + 1.425\log L_R$. {We note that }\citet{West2009} {found the} \citet{Bell2003}{ conversion to be biased by emission lines within the SDSS broadband filters, particularly for the bluest galaxies. However, the EWs in our sample are low (Fig}~\ref{ew50_0_t}{) compared with the $\sim 1000$ \AA $ R$-band filter, so the corrections are small and the conversion is adequate for our purposes.}

In terms of both sSFR and group SFE$_T$, Choir galaxies fall neatly on the best fits to the control SINGG sample, with a KS p-value of 0.37 and 0.14 respectively\footnote{In this section we perform the KS test on the distribution of $\{y - (a + bx)\}$, where $y$ is the sSFR or SFE, $x$ is the stellar mass of the Choir (SINGG) galaxies, and $a$ and $b$ are parameters from the fit to single galaxies in SINGG (galaxies in \citet{Schiminovich2010}). {For the SINGG-Schiminovich comparison we perform the KS test on the subset of stellar masses within the domain of the }\cite{Schiminovich2010} sample.}. This seems in contrast to previous findings that star formation is suppressed at group densities \citep{Lewis2002,Gomez2003}.  However, our selection is different in that typical group catalogues have at least four similarly large galaxies, and are insensitive to the dwarf members. Moreover, our control sample does not consist solely of isolated galaxies; as discussed earlier, at least 20\% of the sample detections are likely to be in similarly dense groups of four or more member {galaxies}. 

We therefore compare the star formation activity for our control sample to the work by \citet[GALEX Arecibo SDSS survey, GASS,][]{Schiminovich2010} and \citet[Arecibo Legacy Fast ALFA  (ALFALFA) survey with SDSS and GALEX photometry,][]{Huang2012}.

Firstly, our control sample exhibits a lower sSFR (by $\sim$ 1 dex across the {corresponding} stellar mass range) than the high-sSFR trend of \citet{Schiminovich2010}, with a KS test p-value $<0.001$. This agrees with our suggestion that many of the galaxies in SINGG are not field galaxies but instead exist in Choir-like groups. Our SINGG sample is more consistent with the ($M_*>9.5$) sSFR trend in \citet{Huang2012}, although their sample shows a much steeper slope than ours. {The SINGG data also hint at a transition to lower sSFR above a turnover stellar mass as seen in }\citet{Bothwell2009} {, but not convincingly so.}

Next, our SFE$_T$ plot (Figure~\ref{MstarVSFRonlogmhi}) is for \emph{groups}, not individual galaxies, but according to \citet{Rownd1999} there should be no variation in SFE with environment. On this basis we compare our SFE data in Figure~\ref{MstarVSFRonlogmhi} to \citet{Schiminovich2010} and \citet{Huang2012}. Our SFE for all of SINGG is lower than the high-sSFR (log SFR/M$_{star}>-11.5$) \citet{Schiminovich2010} data within the corresponding stellar mass range, with a KS-test p-value $<0.001$. Our sample shows increasing SFE with stellar mass, in contrast with the  \citet{Schiminovich2010} data, which do not seem to show any trend. We note that SINGG covers a much wider stellar mass range than the \citet{Schiminovich2010} sample, which may make the small trend more apparent in our work. Our results are more consistent with \citet{Bothwell2009} who found that gas cycling time ($\propto$\:SFE$^{-1}$) decreases shallowly with luminosity (that is, SFE increases slowly with luminosity) for \HI-selected galaxies. Similarly, the SFE work by \citet{Huang2012} also is consistent with our SINGG sample.

We consider the source of discrepancy between our results and those of \citet{Schiminovich2010} and \citet{Huang2012}. Neither we nor \citet{Schiminovich2010} correct for helium content when calculating sSFR or SFE  but both correct for dust absorption. Both assume a \citet{Chabrier2003} IMF. We point out that our SFRs are calculated from H$\alpha$ emission, while the \citet{Schiminovich2010} SFRs are calculated from UV measurements. These indicators for star formation are sensitive to different types of stars; H$\alpha$ probes the formation of the most massive stars ($M_\star > 20 M_\odot$) which have lifetimes $<$ 7 Myr, while UV traces the formation of stars down to $\sim 3 M_\odot$ which have lifetimes up to 300 Myr \citep{Meurer2009}. We converted the NUV-based SFR calibration used by \citet{Schiminovich2010} into the H$\alpha$-based calibration of \citet{Meurer2009} and found that our calibration should yield SFRs 0.2 dex {lower} than \citet{Schiminovich2010} - that is, in the opposite direction to the displayed discrepancy. 

While our sample is selected by \HI\ mass, the \citet{Schiminovich2010} sample has a UV flux-limited selection, biasing their sample towards higher UV-SFRs. The higher redshift range (z$<$0.05) and consequent larger volume of their sample also allows a higher average HI mass and SFR. {Similarly, the} \citet{Huang2012} {sample is also a flux-limited, \HI-selected sample with a higher redshift than SINGG. The brighter and highest-redshift bins have a steep sSFR slope due to the flux limit, while nearby, volume-limited bins have a shallower slope. The combination of these two extremes results in the apparent turnover in their relation (Drinkwater et al., in prep).} The difference between our sample and  \citet{Huang2012} also includes different algorithms for calculating $M_\star$ and SFR to those we use. They use spectral energy density (SED) fitting to get both these quantities, and note that the $M_\star$ estimates are primarily dependent on the reddest fluxes while the SFR estimates come primarily from UV fluxes. We conclude that the differences between our results and those of \citet{Huang2012} and \citet{Schiminovich2010} {are due to differences in sample selection and the calibration of the quantities involved}.

\subsection{\HI\ Deficiency\label{HIDef}}

In general, galaxies in high density environments such as galaxy clusters and groups have less \HI\ than galaxies of the same size and luminosity residing in the field \citep{Haynes1983,Solanes2001,Kilborn2009}. 
This deficiency in \HI\ is quantified by the \HI\ deficiency parameter, defined as the difference between the logarithms of the expected (M$_{HI exp}$) and observed  \HI\ mass (M$_{HI obs}$) of a galaxy \citep{Haynes1983}:
$$\textsc{Def}_{HI} = \text{log[M}_{HI exp}] - \text{log[M}_{HI obs}].$$  

An \HI\ deficiency parameter of 0.3 dex translates into half the \HI\ mass that we would expect a galaxy to have based on its optical luminosity or size. We consider an \HI\ deficiency between -0.3 and 0.3 as normal \HI\ content, {as per }\citet{Kilborn2009}. In this section we exclude HIPASS J0205-55 due to the two HIPASS detections (see Appendix A), and HIPASS J2318-42a because one member is not completely within our field of view.

We used two independent methods to calculate the expected \HI\ content for the Choir group galaxies. {Our first method is to use} the \HI\ scaling relation in D\'{e}nes et al. (2013, in prep.). {This relation is found from an analysis of the HOPCAT }\citep[\HIPASS\ optical catalogue,][]{Doyle2005} and gives \HI\ {mass (M$_{HI}$) as a function of SuperCosmos $R$-band magnitude (Mag$_{R_{SC}}$):}
$$\text{log(M}_{HI}) = 3.82 - 0.3\text{Mag}_{R_{SC}}$$
{We compared the SuperCosmos $R$-band magnitudes in HOPCAT  to our SINGG $R$-band (AB) magnitudes (Mag$_{R_{AB}}$) and found them to scale by Mag$_{R_{SC}}$ = 8.7 + 1.36Mag$_{R_{AB}}$.}

\begin{figure}
\centerline{
\includegraphics[width=1\linewidth]{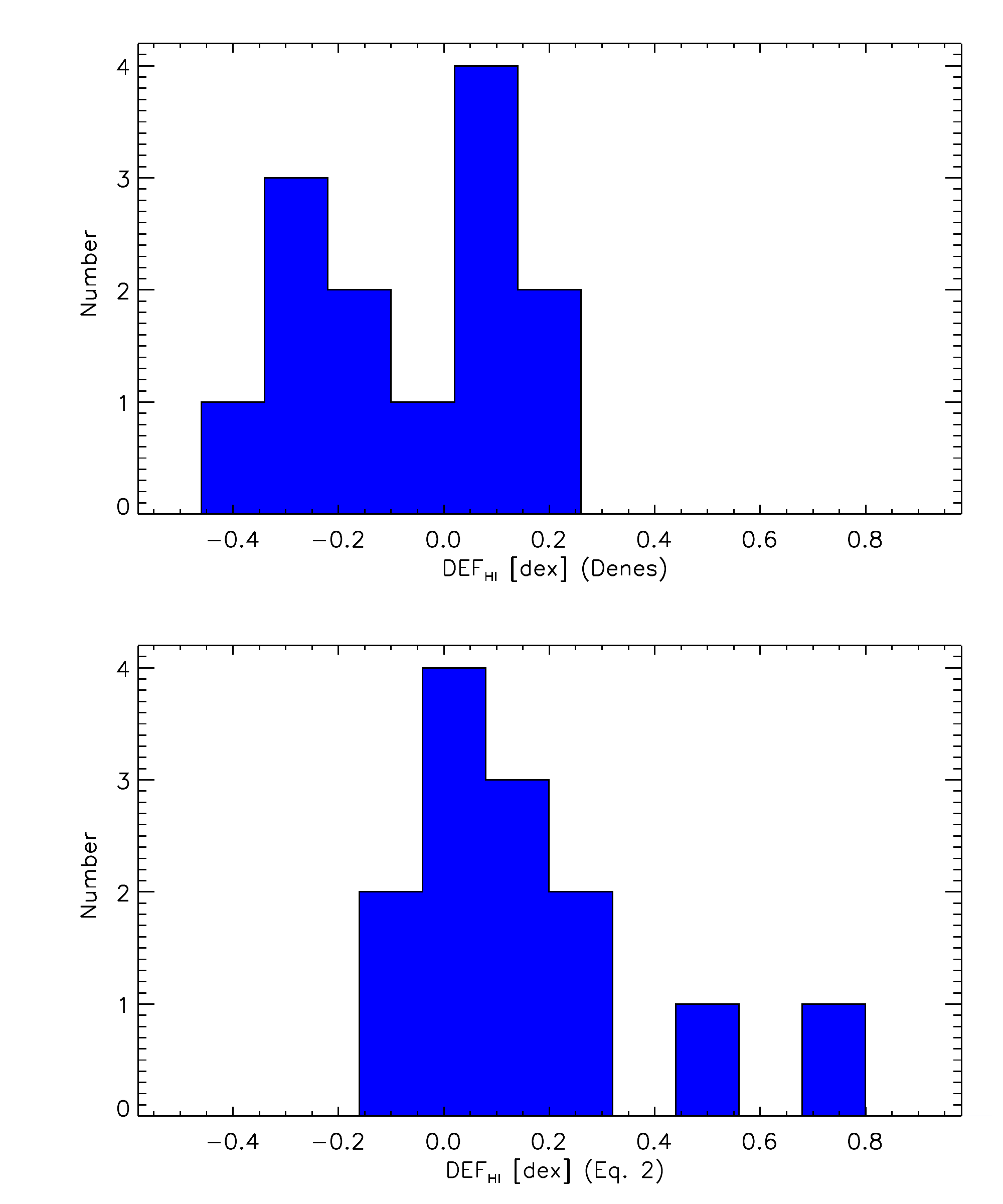}
}
\caption{Distribution of \HI\ deficiency parameter DEF$_{HI}$ for each Choir group, defined as the logarithmic difference between observed group \HI\ mass and predicted group \HI\ mass { (determined by summing the predicted} \HI\ {masses of the individual group  galaxies)}. Our groups are on average not significantly \HI\ deficient. Upper panel: expected \HI\ mass based on $R$-band magnitude. Lower panel: expected \HI\ masses based on our Equation 2. Two very H$\alpha$-luminous groups, HIPASS J1059-09 and J1403-06 are not significantly \HI\ deficient in this definition.
\label{deficiency}}
\end{figure}

The inherent scatter in this relation is $\pm$ 0.3 dex. We then summed over all the members in each group and compared this to the measured \HI\ content to calculate the total \HI\ deficiency for each group. Our results are presented in Figure~\ref{deficiency} (upper panel). 

{Our second method for calculating the expected} \HI\ {content is to use} Equation 2 from this paper, which gives \HI\ mass based on H$\alpha$ luminosity and $R$-band surface brightness. This is shown in the lower panel of Figure~\ref{deficiency}. Again, nearly all of our groups have normal \HI\ content, with the exception of \HIPASS\ J1059-09 and J1403-06, the two groups with the highest H$\alpha$ luminosity in our sample. The {members of the}se two groups also have a high surface brightness, resulting in the highest total predicted \HI\ mass in our sample. {In fact, the two to three brightest members in both groups all have a higher predicted} \HI\ {mass than the corresponding groups themselves (see Fig}~\ref{distanceVlogmhirain}). {The uncertainty in the }\HI\ { mass measurements of $\sim$ 10\% }\citep{Koribalski2004}{, or 0.04 dex is negligible compared with the inherent scatter in Equation 2 of 0.48, so we adopt 0.48 dex as the uncertainty in} \HI\ {deficiency. Hence the deficiency of these two groups is not statistically significant in our definition.}

 The two different methods produce {slightly} different results because the scaling is based on different physical properties. That is, method (1) identifies groups as \HI\ deficient when their stellar luminosity is high compared to their \HI\ mass, while deficient groups in method (2) have a high SFR for their \HI\ mass. The implication is that the two groups that are deficient by method (2) and not (1) are dominated by high H$\alpha$ equivalent width starbursting galaxies.
 
{The fact that the Choir groups show no significant }\HI\ { deficiency is a similar result to  }\citet{Kilborn2009} {who showed an average lack of }\HI\ { deficiency for their sample of optically-selected loose galaxy groups. The situation is less clear for compact groups, with }\citet{Stevens2004} {finding no significant }\HI\ {deficiency, while }\citet{Borthakur2010}{, found the typical }\HI\ {deficiency of their sample of Hickson compact groups to be between 0.2-0.4 dex; in several cases the deficiency exceeded 0.5 dex.}

We also compare the Choir groups to the gas-rich M81 group, as modelled by \citet{Nichols2011} in order to explain the \HI\ deficiency of the LG \citep{Grcevich2009}. They found that the M81 group must have commenced assembly at  $z\sim2$, in contrast to the LG which must have started by $z\sim10$. The overall lack of \HI\ deficiency of the Choir groups suggests that the group environment has not yet removed substantial amounts of \HI\ gas from these groups.  Hence the Choir groups are at an early stage of assembly.  In the local context, this would make them more like the M81 group than the LG. Consequently we expect that, like the M81 group, the Choir groups have a larger system of HI clouds than the LG does. The fact that the Choir groups are gas-rich and less evolved than the Local Group indicates that they may provide important information about how gas enters groups and galaxies. 

\section{Conclusions}
In this paper we have presented the Choirs: fields of four or more H$\alpha$-emitting galaxies found in the Survey for Ionisation for Neutral Gas Galaxies (SINGG). We found fifteen such groups in SINGG. 

We make the following points:
\begin{enumerate}
\item Due to selection effects, Choirs groups are at the large distance, high mass end of the parent SINGG sample of \HI\ sources.
\item Choir member galaxies are not significantly different from the control SINGG sample in any of our measures of radius, H$\alpha$ equivalent width, $R$-band surface brightness, specific star formation rate or star formation efficiency. 
\item The dwarf galaxies in our Choir groups are not detectable on their own in \HIPASS, but are detected in SINGG because the entire group has sufficient \HI\ to be selected in \HIPASS.
\item Within the limitations of the SINGG imaging field of view, there are no giant elliptical galaxies in the Choir groups.
\item Eight of the fifteen Choir groups are characterised by having two giant spiral galaxies and a number of smaller galaxies. In terms of {\emph{morphology}} they can be considered to be Local Group analogues.
\item The mean group projected size is very compact at 190 kpc; much smaller than groups in the \citet{Garcia1993} catalogue at 961 kpc, although not as compact as \citet{Hickson1989} Compact Groups at 87 kpc. The mean Choir compactness is also more than 3$\sigma$ smaller than the same statistic for the Local Group. We note that our group size is limited by the field of view, with a maximum size of 380 kpc at the mean distance of 87 Mpc.
\item The specific star formation rate (sSFR = SFR/$M_\star$) of Choir member galaxies falls on the same $M_\star$ scaling relation as the rest of SINGG.  This scaling relation is similar to what is found by for the 
ALFALFA HI selected survey \citep{Huang2012}.  However, galaxies from the $M_\star$-selected GASS survey \citep{Schiminovich2010} have sSFR 0.5 dex higher than our sample. Differences in the selection of the different samples, the depth of the observations, and the SFR calibrations are likely to account for the differences between these surveys.
\item The star formation efficiency (SFE = SFR/$M_{HI}$) of the Choir groups matches the sample of remaining SINGG members, which in turn is lower than the portion of the \citet{Schiminovich2010} sample with high sSFR. Our SINGG sample shows an increasing trend in SFE with stellar mass, consistent with \citet{Bothwell2009} and \citet{Huang2012}.
\item On average our groups are not significantly \HI\ deficient, unlike typical groups of galaxies. This suggests an earlier stage of assembly than the Local Group, and more like the M81 group \citep{Nichols2011}.
\item Our results indicate that emission line selection is an efficient way to pick out {candidate} galaxy groups in blind \HI\ surveys. This can be very important when the beam size is large compared to the separations of galaxies within groups. Here, it is the H$\alpha$ imaging that allows the small ELGs to be identified as likely dwarf group members. In comparison, astronomers using UV imaging alone to identify ELGs \citep[e.g.][]{Huang2012} may be reluctant to identify the smaller sources as dwarf members without follow-up spectroscopy. 
\end{enumerate}

In summary, \HI\ combined with H$\alpha$ selection can result in the selection of \HI-rich groups.  These are fairly compact and typically contain sources with strong signs of interaction, although global properties appear fairly normal.  {In approximately half of the cases}, the groups are similar to the Local Group in containing two bright large spirals and numerous dwarf galaxies, although the compactness suggests the groups are at a more advanced stage of interaction than the LG. The lack of \HI\ deficiency suggests that the groups are at an earlier stage of group assembly, more like the M81 group.

\section*{Acknowledgements}
The authors wish to thank the anonymous referee for their helpful comments, David Schiminovich for making his data available to us, and Aaron Robotham for useful discussions.

We acknowledge funding support from the University of Queensland - University of Western Australia Bilateral Research Collaboration Award and the Australian Research Council (grant DP110102608).

This research has made use of the NASA's Astrophysics Data System.

We also made use of the NASA/IPAC Extragalactic Database (NED) which is operated by the Jet Propulsion Laboratory, California Institute of Technology, under contract with the National Aeronautics and Space Administration.

This research used the facilities of the Canadian Astronomy Data Centre operated by the National Research Council of Canada with the support of the Canadian Space Agency.

J.H.K. is supported by Korean Research Foundation (KRF) grant funded by the Korean government (MEST), No. 2010--0000712.

\addcontentsline{toc}{chapter}{Bibliography}
\bibliographystyle{hapj}
\bibliography{Choirs1Catalogue}
\clearpage

\onecolumn
\begin{landscape}
\begin{table*}
\caption{Summary of Choir groups\label{groups}}
\begin{centering}
\begin{tabular}{|l|l|l|l|l|l|l|l|l|l|l|l|l|l|l|}
\hline
HIPASS+  & Optical ID & RA          & Dec          & Dist. & FOV    & ELGs & Comp. & M$_{HI}$& \HI\ Def.& V$_{HI}$ & $W_{HI}$& $W_{50,F}$    \\        
         &            &  [h m s]    &  [d m s]     & [Mpc] & [kpc]  &      & [kpc] & [dex]   & [dex]   & [km/s]   & [km/s]  & [km/s]        \\      
(1)      & (2)        & (3)         & (4)          & (5)   & (6)    & (7)  & (8)   & (9)     & (10)    & (11)     & (12)    & (13)          \\ \hline 
J0205-55 & AM 0203-552& 02 05 05.48 & -55 06 42.55 & 93    & 406    & 9    & 366   & 10.51   & 0.03**  & 6524     & 193     & 5051-8297     \\  
J0209-10 & HCG 16     & 02 09 42.71 & -10 11 01.36 & 54    & 236    & 4    & 56.2  & 10.31   & 0.18    & 3900     & 243     & 2560-5668     \\      
J0258-74 &            & 02 58 06.48 & -74 27 22.79 & 70    & 305    & 4    & 236   & 10.41   & -0.34   & 4805     & 399     & 2560-5668     \\      
J0400-52 & Abell 3193 & 04 00 40.82 & -52 44 02.72 & 151   & 659    & 9    & 420   & 10.61   & 0.13    & 10566    & 298     & 8182-11750    \\     
J0443-05 &            & 04 43 43.90 & -05 19 09.91 & 69    & 301    & 5    & 209   & 10.41*  & 0.02    & 4877     & 278     & 2560-5668     \\      
J1026-19 &            & 10 26 40.81 & -19 03 04.03 & 135   & 589    & 6    & 107   & 10.63   & -0.28   & 9094     & 242     & 6857-9142***  \\    
J1051-17 &            & 10 51 37.46 & -17 07 29.24 & 83    & 362    & 9    & 216   & 10.45   & -0.26   & 5477     & 522     & 4205-7679     \\      
J1059-09 & USGC S154  & 10 59 16.25 & -09 47 38.15 & 122   & 532    & 10   & 283   & 10.42*  & 0.20    & 8175     & 80.0    & 6857-9142     \\  
J1159-19 & ARP 022    & 11 59 30.13 & -19 15 54.86 & 25    & 109    & 4    & 29.5  & 9.92    & 0.00    & 1668     & 150     & 1188-2651     \\      
J1250-20 &            & 12 50 52.84 & -20 22 15.64 & 114   & 497    & 7    & 123   & 10.51   & -0.11   & 7742     & 169     & 5051-8297     \\      
J1403-06 & ARP 271    & 14 03 24.88 & -06 04 09.16 & 41    & 179    & 4    & 27.5  & 10.29   & 0.09    & 2591     & 330     & 2217-3725     \\      
J1408-21 &            & 14 08 42.04 & -21 35 49.81 & 128   & 559    & 6    & 184   & 10.52   & 0.05    & 8732     & 203     & 6857-9142     \\      
J1956-50 &            & 19 56 45.51 & -50 03 20.30 & 110   & 480    & 4    & 299   & 10.52   & -0.33   & 7446     & 321     & 4205-7679***  \\    
J2027-51 & AM 2024-515& 20 28 06.39 & -51 41 29.83 & 87    & 380    & 4    & 224   & 10.44   & -0.22   & 5881     & 356     & 4205-7679     \\      
J2318-42a&Grus Quartet& 23 16 10.80 & -42 35 05.00 & 23    & 100    & 4    & 73.1  & 10.10   & 0.15**  & 1575     & 222     & 1188-2651     \\      

\hline
\end{tabular}
\end{centering}
\\\justifying{Columns: (1): name assigned to field in \HIPASS; (2): name assigned to group as found in NASA/IPAC Extragalactic Database (NED; http://ned.ipac.caltech.edu/); (3): J2000 right ascension of brightest source in field; (4): J2000 declination of brightest source in field; (5): distance based on the multipole attractor model as in \citet{Mould2000} and adopting H$_0$ = 70 km s$^{-1}$ Mpc$^{-1}$; (6) field of view of $\sim$15 arcminutes at the distance of the group; (7): number of ELGs; (8): projected group compactness, estimated by projected separation of two largest group members. See Section~\ref{compactness}.; (9): logarithm of group \HI\ mass from \HIPASS. (*) We remeasured \HI\ mass for two groups whose \HIPASS\ measurements were incorrect. See Section 2 for more details.; (10): \HI\ deficiency parameter {defined as the logarithmic difference between the observed group} \HI\ {mass and predicted group} \HI\ {mass (determined by summing the predicted} \HI\ {masses of the individual group galaxies)}. See Section~\ref{HIDef} for calculation and discussion. (**) We exclude these two groups from our \HI\ deficiency analysis due to field of view restrictions.;  (11): heliocentric \HI\ velocity; (12): observed \HI\ emission width; (13): narrow-band filter velocity range. {(***) We note that the narrow-band filters used for these two fields overlap but do not completely cover the extent of the observed} \HI\ {emission width. Therefore there may be additional ELGs associated with these groups which would be classified as Choir member galaxies but are not detected in our imaging.;}}
\end{table*}
\end{landscape}

\begin{landscape}
\setlength{\LTcapwidth}{25cm}
\begin{longtable}{|l|l|l|l|l|l|l|l|l|l|l|}
\multicolumn{11}{c}{{\bfseries \tablename:  Catalogue of Choir members}} \\
\hline
\multicolumn{1}{|l|}{SINGG name} &\multicolumn{1}{|l|}{Optical ID}&\multicolumn{1}{|l|}{RA} &\multicolumn{1}{|l|}{Dec}&\multicolumn{1}{|l|}{$r_e$} &\multicolumn{1}{|l|}{Axial}&\multicolumn{1}{|l|}{M$_R$}&\multicolumn{1}{|l|}{log(F$_{H\alpha}$)}&\multicolumn{1}{|l|}{$\mu_e$}&\multicolumn{1}{|l|}{Morphology}&\multicolumn{1}{|l|}{Radial velocity}\\
\multicolumn{1}{|l|}{H{\sc i}PASS+} &\multicolumn{1}{|l|}{}&\multicolumn{1}{|l|}{[h m s]} &\multicolumn{1}{|l|}{[d m s]}&\multicolumn{1}{|l|}{[arcsec] } &\multicolumn{1}{|l|}{ratio}&\multicolumn{1}{|l|}{[ABmag]}&\multicolumn{1}{|l|}{[erg/s/cm$^2$]}&\multicolumn{1}{|l|}{[ABmag/''$^2$]}&\multicolumn{1}{|l|}{}&\multicolumn{1}{|l|}{[km/s]}\\
\multicolumn{1}{|l|}{(1)} &\multicolumn{1}{|l|}{(2)}&\multicolumn{1}{|l|}{(3)} &\multicolumn{1}{|l|}{(4)}&\multicolumn{1}{|l|}{(5) } &\multicolumn{1}{|l|}{(6)}&\multicolumn{1}{|l|}{(7)}&\multicolumn{1}{|l|}{(8)}&\multicolumn{1}{|l|}{(9)}&\multicolumn{1}{|l|}{(10)}&\multicolumn{1}{|l|}{(11)}
\\
\hline
\hline
\endfirsthead
\multicolumn{11}{c}{{\bfseries \tablename : continued from previous page}} \\*
\hline
\multicolumn{1}{|l|}{SINGG name} &\multicolumn{1}{|l|}{Optical ID}&\multicolumn{1}{|l|}{RA} &\multicolumn{1}{|l|}{Dec}&\multicolumn{1}{|l|}{$r_e$} &\multicolumn{1}{|l|}{Axial}&\multicolumn{1}{|l|}{M$_R$}&\multicolumn{1}{|l|}{log(F$_{H\alpha}$)}&\multicolumn{1}{|l|}{$\mu_e$}&\multicolumn{1}{|l|}{Morphology}&\multicolumn{1}{|l|}{Radial velocity}\\
\multicolumn{1}{|l|}{H{\sc i}PASS+} &\multicolumn{1}{|l|}{}&\multicolumn{1}{|l|}{[h m s]} &\multicolumn{1}{|l|}{[d m s]}&\multicolumn{1}{|l|}{[arcsec] } &\multicolumn{1}{|l|}{ratio}&\multicolumn{1}{|l|}{[ABmag]}&\multicolumn{1}{|l|}{[erg/s/cm$^2$]}&\multicolumn{1}{|l|}{[ABmag/''$^2$]}&\multicolumn{1}{|l|}{}&\multicolumn{1}{|l|}{[km/s]}\\
\multicolumn{1}{|l|}{(1)} &\multicolumn{1}{|l|}{(2)}&\multicolumn{1}{|l|}{(3)} &\multicolumn{1}{|l|}{(4)}&\multicolumn{1}{|l|}{(5) } &\multicolumn{1}{|l|}{(6)}&\multicolumn{1}{|l|}{(7)}&\multicolumn{1}{|l|}{(8)}&\multicolumn{1}{|l|}{(9)}&\multicolumn{1}{|l|}{(10)}&\multicolumn{1}{|l|}{(11)}
\\
\hline
\hline
\endhead
\hline
\hline
\caption{Choir member descriptions.\label{choirs}\newline Columns: (1): Name assigned in SINGG (H{\sc i}PASS name with :S\emph{n} appended for \emph{n}th source;  (2) Previously assigned ID based on position match with NASA/IPAC Extragalactic Database (NED; http://ned.ipac.caltech.edu/); (3): Right Ascension (J2000); (4): Declination (J2000); (5): R-band effective radius; (6): Ratio of major and minor axes; (7): R-band absolute magnitude; (8): Logarithm of total H$\alpha$ flux, corrected for Galactic extinction and [N{\sc ii}] contamination; (9): R-band effective surface brightness, face-on and extinction-corrected (Galactic and internal); (10): NED morphological classification where available, or [new classification]; (11): Central radial velocity from our ANU2.3mT/WiFeS data (Sweet et al., in prep). 
Sources marked with * have been identified since SINGG release 1; for these we give preliminary  measurements performed using IRAF's {\sc apphot} task. IRAF is distributed by the National Optical Astronomy Observatories, which are operated by the Association of Universities for Research in Astronomy, Inc., under cooperative agreement with the National Science Foundation (Tody, 1993).
{**} Our declination for J1250-20:S3 is 10'' different to the ESO measurement, suggesting a problem with their search radius for this object.
{***} Only half of J2318-42a is within our FOV, so we use the NED position for J2318-42a:S3.}\\
\hline
\multicolumn{11}{|r|}{{Continued on next page}} \\ 
\hline
\endfoot
\hline 
\hline
\endlastfoot
J0205-55:S1         & ESO153-G017
                    & 02 05 05.48 & -55 06 42.54 & 17.60$\pm$0.22 & 1.66 & -21.95$\pm$0.22 & -12.01$\pm$0.14 & 20.43$\pm$0.02 &SAB(r)bc&6491\\
J0205-55:S2         & ESO153-IG016                   & 02 04 50.78 & -55 13 01.55 & 05.19$\pm$0.00 & 2.26 & -18.91$\pm$0.00 & -12.86$\pm$0.04 & 21.12$\pm$0.02 &SB(s)cd pec&5942\\
J0205-55:S3         & ESO153-G015                    & 02 04 34.92 & -55 07 09.65 & 06.72$\pm$0.02 & 1.92 & -21.35$\pm$0.02 & -12.94$\pm$0.70 & 18.94$\pm$0.02 &S0&\\
J0205-55:S4         & ESO153-G013                    & 02 04 19.75 & -55 13 50.44 & 11.25$\pm$0.09 & 3.45 & -21.39$\pm$0.09 & -12.61$\pm$0.34 & 20.02$\pm$0.02 &Sa:&5942\\
J0205-55:S5         & APMUKS
& 02 04 54.77 & -55 08 31.99 & 05.84$\pm$0.18 & 2.19 & -17.13$\pm$0.18 & -14.45$\pm$0.36 & 23.28$\pm$0.02 &[D]&6127\\
J0205-55:S6         & APMUKS
& 02 04 57.07 & -55 13 34.10 & 02.52$\pm$0.06 & 1.56 & -18.14$\pm$0.06 & -14.13$\pm$0.36 & 20.38$\pm$0.03 &[D]&5760\\
J0205-55:S7         & 6dF
& 02 05 00.57 & -55 15 19.63 & 01.87$\pm$1.08 & 1.54 & -15.65$\pm$1.08 & -14.35$\pm$0.29 & 22.35$\pm$0.35 &[cD]&5760\\
J0205-55:S8         & APMUKS
& 02 04 29.71 & -55 12 56.09 & 02.61$\pm$0.05 & 1.03 & -17.44$\pm$0.05 & -14.69$\pm$0.69 & 21.21$\pm$0.02 &[cD]&\\
J0205-55:S9*	     & APMUKS
& 02 05 23.76 & -55 14 14.20 & 01.06$\pm$0.21  & --   & -15.37$\pm$0.12 & -14.39$\pm$0.46  & 18.78$\pm$0.45&[cD]&\\
J0209-10:S1          & NGC0839                        & 02 09 42.71 & -10 11 01.36 & 11.93$\pm$0.12 & 2.32 & -21.00$\pm$0.12 & -12.11$\pm$0.11 & 19.44$\pm$0.02 &Spec sp; LINER Sy2&\\
J0209-10:S2          & NGC0838                        & 02 09 38.48 & -10 08 45.79 & 07.97$\pm$0.07 & 1.33 & -21.16$\pm$0.07 & -11.38$\pm$0.03 & 18.36$\pm$0.02 &SA(rs)0$^{\circ}$ pec: Sbrst&\\
J0209-10:S3          & NGC0835                        & 02 09 24.43 & -10 08 10.59 & 16.04$\pm$0.18 & 2.25 & -21.47$\pm$0.18 & -11.90$\pm$0.12 & 19.53$\pm$0.03 &SAB(r)ab: pec LINER&\\
J0209-10:S4          & NGC0833                        & 02 09 20.69 & -10 07 58.55 & 08.87$\pm$0.01 & 1.73 & -21.14$\pm$0.01 & -12.68$\pm$0.49 & 18.63$\pm$0.02 &(R')SA:pec;Sy2 LINER&\\
J0258-74:S1          & ESO031-G005                    & 02 58 06.48 & -74 27 22.79 & 20.25$\pm$0.26 & 2.81 & -21.50$\pm$0.26 & -12.06$\pm$0.09 & 20.47$\pm$0.02 &SAB(rs)bc HII&\\
J0258-74:S2          & MRSS
& 02 58 52.43 & -74 25 53.25 & 10.34$\pm$0.29 & 1.22 & -19.15$\pm$0.29 & -13.04$\pm$0.08 & 21.66$\pm$0.03 &[S]&\\
J0258-74:S3          & 2MASX
& 02 58 42.76 & -74 26 03.55 & 06.72$\pm$0.56 & 3.40 & -18.25$\pm$0.56 & -13.47$\pm$0.12 & 21.70$\pm$0.09 &[S]&\\
J0258-74:S4          & MRSS
& 02 57 29.23 & -74 22 34.75 & 04.07$\pm$0.85 & 1.31 & -17.13$\pm$0.85 & -14.05$\pm$0.33 & 21.80$\pm$0.27 &[dIrr]&\\
J0400-52:S1          & ESO156-G029                    & 04 00 40.82 & -52 44 02.71 & 06.80$\pm$0.06 & 1.22 & -21.32$\pm$0.06 & -12.84$\pm$0.20 & 20.11$\pm$0.02 &SA(rs)cd pec:&\\ 
J0400-52:S2          & APMUKS
& 04 00 48.07 & -52 41 02.81 & 01.75$\pm$0.02 & 1.44 & -15.78$\pm$0.02 & -14.67$\pm$0.10 & 23.15$\pm$0.02 &[S]&\\
J0400-52:S3          & 2MASX   
& 04 00 06.03 & -52 39 32.63 & 02.95$\pm$0.01 & 1.64 & -19.64$\pm$0.01 & -13.61$\pm$0.20 & 20.18$\pm$0.02 &[S]&\\
J0400-52:S4          & IC2028 			      & 04 01 18.23 & -52 42 27.08 & 08.10$\pm$0.01 & 1.47 & -22.07$\pm$0.01 & -12.63$\pm$0.31 & 19.57$\pm$0.02 &Scd:&\\
J0400-52:S5          & 2MASX
& 04 00 53.00 & -52 49 38.43 & 09.41$\pm$0.05 & 1.42 & -21.98$\pm$0.05 & -12.65$\pm$0.30 & 20.01$\pm$0.02 &(R)SB(s)b? pec&\\
J0400-52:S6          & IC2029			      & 04 01 17.84 & -52 48 02.81 & 12.08$\pm$0.06 & 1.80 & -21.50$\pm$0.06 & -13.05$\pm$0.42 & 21.13$\pm$0.02 &SB(s)c pec&\\
J0400-52:S7          & APMUKS
& 04 01 08.99 & -52 49 32.78 & 03.72$\pm$0.13 & 1.59 & -18.61$\pm$0.13 & -14.08$\pm$0.23 & 21.81$\pm$0.04 &[D]&\\
J0400-52:S8*	     & -			      & 04 01 17.00 & -52 42 08.50 & 02.07$\pm$0.35 & --   & -17.36$\pm$0.08 & -14.20$\pm$0.52 & 16.46$\pm$0.37 &[D]&\\
J0400-52:S9*	     & -			      & 04 01 19.29 & -52 47 56.10 & 03.46$\pm$1.38  & --   & -17.56$\pm$0.07 & -14.09$\pm$0.57 & 15.15$\pm$0.87 &[D]&\\
J0443-05:S1          & NGC1643                        & 04 43 43.90 & -05 19 09.93 & 10.71$\pm$0.21 & 1.19 & -21.52$\pm$0.21 & -11.65$\pm$0.04 & 19.05$\pm$0.04 &SB(r)bc pec?&\\
J0443-05:S2          & NGC1645                        & 04 44 06.43 & -05 27 56.31 & 16.90$\pm$0.15 & 1.99 & -21.87$\pm$0.15 & -12.15$\pm$0.20 & 19.62$\pm$0.02 &(R')SB(rs)0+ pec&\\
J0443-05:S3          & 2MASX
& 04 44 11.67 & -05 14 38.31 & 04.79$\pm$0.19 & 1.86 & -19.56$\pm$0.19 & -13.38$\pm$0.25 & 19.53$\pm$0.06 &[S]&\\
J0443-05:S4          & 2MASX
& 04 44 05.54 & -05 25 46.50 & 04.95$\pm$0.12 & 1.60 & -18.77$\pm$0.12 & -13.07$\pm$0.06 & 20.46$\pm$0.03 &[D]&\\ 
J0443-05:S5*	     & SEGC
& 04 43 45.02 & -05 19 41.90 & 01.97$\pm$0.98  & --   & -17.65$\pm$0.04	 & -13.47$\pm$0.91 & 14.62$\pm$1.09 &S0+? pec&\\
J1026-19:S1          & ESO568-G011                    & 10 26 40.81 & -19 03 04.01 & 11.86$\pm$0.19 & 1.28 & -22.01$\pm$0.19 & -12.49$\pm$0.35 & 20.12$\pm$0.03 &SAB(s)bc: pec Sbrst&\\
J1026-19:S2          & 2MASX
& 10 26 50.07 & -19 04 31.77 & 05.46$\pm$0.20 & 1.36 & -20.16$\pm$0.20 & -13.04$\pm$0.15 & 20.57$\pm$0.05 &Irr&\\
J1026-19:S3          & -                       & 10 26 18.93 & -18 57 52.12 & 04.68$\pm$0.27 & 1.45 & -18.41$\pm$0.27 & -14.36$\pm$0.54 & 22.16$\pm$0.03 &[D]&\\
J1026-19:S4          & -                       & 10 26 24.40 & -19 02 02.99 & 02.19$\pm$0.26 & 1.38 & -17.62$\pm$0.26 & -14.26$\pm$0.35 & 21.35$\pm$0.11 &[dIrr]&\\
J1026-19:S5          & -                       & 10 26 42.07 & -19 07 35.07 & 05.56$\pm$0.60 & 1.55 & -16.87$\pm$0.60 & -14.67$\pm$0.42 & 24.17$\pm$0.04 &[dIrr]&\\
J1026-19:S6          & FLASH
& 10 26 25.21 & -19 10 35.31 & 07.91$\pm$0.07 & 1.10 & -19.83$\pm$0.07 & -14.27$\pm$1.67 & 21.75$\pm$0.02 &S0&\\
J1051-17:S1          & 2MASX
& 10 51 37.45 & -17 07 29.23 & 27.62$\pm$0.26 & 1.76 & -21.01$\pm$0.26 & -12.62$\pm$0.26 & 22.12$\pm$0.02 &(R?+PR?)&5485\\
J1051-17:S2          & NGC3431                        & 10 51 15.11 & -17 00 29.44 & 14.72$\pm$0.05 & 2.47 & -21.25$\pm$0.05 & -12.40$\pm$0.21 & 20.46$\pm$0.02 &SABb?&5302\\
J1051-17:S3          & -                       & 10 51 35.94 & -16 59 16.80 & 06.61$\pm$0.05 & 1.02 & -17.91$\pm$0.05 & -13.72$\pm$0.13 & 22.43$\pm$0.02 &[dS]&5988\\
J1051-17:S4          & -                       & 10 51 26.01 & -17 05 03.61 & 03.48$\pm$0.09 & 1.40 & -16.19$\pm$0.09 & -14.51$\pm$0.19 & 22.84$\pm$0.02 &[dIrr]&5485\\
J1051-17:S5          & -                       & 10 51 50.91 & -16 58 31.64 & 03.58$\pm$0.06 & 1.75 & -17.02$\pm$0.06 & -14.35$\pm$0.33 & 22.03$\pm$0.02 &[dIrr]&5485\\
J1051-17:S6          & -                       & 10 51 42.78 & -17 06 34.59 & 02.11$\pm$0.04 & 1.29 & -16.78$\pm$0.04 & -14.18$\pm$0.13 & 21.14$\pm$0.02 &[cD]&5668\\
J1051-17:S7          & -                       & 10 51 33.36 & -17 08 36.63 & 04.18$\pm$0.12 & 1.53 & -16.77$\pm$0.12 & -14.28$\pm$0.21 & 22.63$\pm$0.02 &[cD]&5394\\
J1051-17:S8*	     & -			      & 10 51 25.92 & -17 08 16.44 & 01.52$\pm$0.67 & --   & -17.48$\pm$0.04 & -13.46$\pm$0.88 & 15.44$\pm$0.97 &[dS]&5314\\ 
J1051-17:S9*	     & -			      & 10 51 56.54 & -17 05 03.50 & 00.34$\pm$0.17 & --   & -16.68$\pm$0.05 & -13.78$\pm$0.71 & 19.51$\pm$1.09 &[dE,N]&5606\\ 
J1059-09:S1          & MCG-01-28-013                  
& 10 59 16.25 & -09 47 38.16 & 15.08$\pm$0.20 & 1.91 & -22.26$\pm$0.20 & -12.12$\pm$0.12 & 20.20$\pm$0.03 &SAB(rs)b pec:&\\ 
J1059-09:S2          & GNX034                         & 10 59 06.77 & -09 45 04.38 & 11.72$\pm$0.24 & 1.36 & -19.80$\pm$0.24 & -12.91$\pm$0.05 & 22.51$\pm$0.02 &[Irr]&\\
J1059-09:S3          & MCG-01-28-012                  
& 10 59 15.61 & -09 48 59.41 & 14.66$\pm$0.37 & 3.13 & -20.84$\pm$0.37 & -12.51$\pm$0.06 & 21.82$\pm$0.03 &Sab pec sp&\\
J1059-09:S4          & MRK1273                        & 10 58 46.84 & -09 50 43.31 & 08.79$\pm$0.15 & 1.32 & -21.53$\pm$0.15 & -12.48$\pm$0.12 & 19.91$\pm$0.03 &SB0-a Sbrst&\\
J1059-09:S5          & GNX066                         & 10 59 30.98 & -09 44 25.26 & 09.11$\pm$0.13 & 2.84 & -19.57$\pm$0.13 & -13.15$\pm$0.08 & 22.21$\pm$0.02 &[S]&\\
J1059-09:S6          & -                       & 10 59 08.46 & -09 43 14.49 & 08.04$\pm$0.16 & 1.69 & -19.00$\pm$0.16 & -13.54$\pm$0.12 & 22.56$\pm$0.02 &[Irr]&\\
J1059-09:S7          & -                       & 10 59 21.31 & -09 47 50.49 & 02.50$\pm$0.15 & 1.59 & -18.83$\pm$0.15 & -13.47$\pm$0.08 & 20.21$\pm$0.06 &[cD]&\\
J1059-09:S8          & -                       & 10 59 01.73 & -09 52 46.76 & 03.46$\pm$0.40 & 1.81 & -16.81$\pm$0.40 & -14.91$\pm$0.41 & 23.07$\pm$0.10 &[D]&\\
J1059-09:S9*	     & -			      & 10 58 44.69 & -09 53 28.60 & 00.59$\pm$0.29 & --   & -13.56$\pm$0.10 & -14.38$\pm$0.43 & 19.58$\pm$1.09 &[dIrr]&\\ 
J1059-09:S10*	     & 2MASX & 10 59 02.64 & -09 53 19.90& 00.37$\pm$0.07 & -- & -17.03$\pm$0.02 & -16.16$\pm$0.44 & 17.13$\pm$0.44 &[dS]&\\ 
J1159-19:S1          & NGC4027                        & 11 59 30.13 & -19 15 54.88 & 34.37$\pm$0.13 & 1.07 & -21.24$\pm$0.13 & -10.97$\pm$0.04 & 19.71$\pm$0.02 &SB(s)dm HII&\\
J1159-19:S2          & NGC4027A                       & 11 59 29.34 & -19 19 59.52 & 16.26$\pm$0.10 & 1.40 & -17.83$\pm$0.10 & -12.83$\pm$0.06 & 21.86$\pm$0.02 &IB(s)m:&\\
J1159-19:S3          & -                       & 11 59 35.79 & -19 19 02.99 & 03.80$\pm$0.29 & 1.00 & -15.18$\pm$0.29 & -14.21$\pm$0.28 & 21.48$\pm$0.06 &[D]&\\
J1159-19:S4          & ISZ108A                        & 11 59 37.88 & -19 19 45.80 & 06.18$\pm$0.36 & 1.34 & -15.77$\pm$0.36 & -14.04$\pm$0.23 & 21.91$\pm$0.05 &[D]&\\
J1250-20:S1          & ESO575-G006                    & 12 50 52.84 & -20 22 15.65 & 09.95$\pm$0.13 & 1.38 & -21.85$\pm$0.13 & -12.17$\pm$0.10 & 19.58$\pm$0.03 &SA(s)bc pec HII&\\
J1250-20:S2          & ESO575-G004                    & 12 50 40.91 & -20 20 06.22 & 10.09$\pm$0.04 & 1.37 & -21.44$\pm$0.04 & -12.44$\pm$0.11 & 20.09$\pm$0.02 &S&\\
J1250-20:S3**          & ESO-LV5750061                 & 12 50 49.77 & -20 22 03.44 & 04.25$\pm$0.14 & 1.54 & -18.57$\pm$0.14 & -13.36$\pm$0.07 & 21.43$\pm$0.04 &[Irr]&\\ 
J1250-20:S4          & -                       & 12 50 39.92 & -20 20 52.87 & 05.93$\pm$0.11 & 1.59 & -18.72$\pm$0.11 & -13.95$\pm$0.20 & 21.98$\pm$0.03 &[Irrs]&\\
J1250-20:S5          & -                       & 12 50 59.16 & -20 28 14.80 & 02.66$\pm$0.27 & 2.04 & -17.43$\pm$0.27 & -13.84$\pm$0.15 & 21.64$\pm$0.10 &[cD]&\\
J1250-20:S6*	     & -			      & 12 50 45.87 & -20 23 30.10 & 0.154$\pm$0.51 & --   & -15.38$\pm$0.15  & -14.39$\pm$0.43 & 18.42$\pm$0.74 &[D]&\\
J1250-20:S7*	     & -			      & 12 50 46.92 & -20 23 12.40 & 01.28$\pm$0.51 & --   & -14.16$\pm$0.27 & -14.83$\pm$0.25 & 20.04$\pm$0.91 &[D]&\\ 
J1403-06:S1          & NGC5426                        & 14 03 24.88 & -06 04 09.14 & 29.64$\pm$0.35 & 1.99 & -21.18$\pm$0.35 & -11.38$\pm$0.03 & 20.57$\pm$0.02 &SA(s)c pec&2512\\
J1403-06:S2          & NGC5427                        & 14 03 26.09 & -06 01 51.20 & 36.45$\pm$0.09 & 1.30 & -22.01$\pm$0.09 & -11.11$\pm$0.04 & 20.05$\pm$0.02 &SA(s)c pec;Sy2 HII&2741\\ 
J1403-06:S3          & APMUKS
& 14 03 13.48 & -06 06 24.17 & 04.18$\pm$0.85 & 1.03 & -15.27$\pm$0.85 & -14.40$\pm$0.26 & 22.70$\pm$0.17 &[cD]&2767\\
J1403-06:S4          & APMUKS
& 14 03 34.62 & -06 07 59.27 & 05.96$\pm$0.86 & 1.43 & -14.42$\pm$0.86 & -14.85$\pm$0.34 & 24.34$\pm$0.02 &[cD]&2685\\
J1408-21:S1          & ESO578-G026                    & 14 08 42.04 & -21 35 49.82 & 10.24$\pm$0.26 & 1.47 & -22.35$\pm$0.26 & -12.24$\pm$0.07 & 19.22$\pm$0.05 &SB(rl)c&8704\\ 
J1408-21:S2          & 2MASX
& 14 08 57.72 & -21 38 52.47 & 07.99$\pm$0.06 & 1.39 & -21.11$\pm$0.06 & -12.67$\pm$0.04 & 20.15$\pm$0.02 &[S]&8831\\
J1408-21:S3          & 2MASX
& 14 08 41.04 & -21 37 40.97 & 06.06$\pm$0.10 & 1.60 & -20.26$\pm$0.10 & -13.19$\pm$0.07 & 20.52$\pm$0.03 &[S0]&8792\\
J1408-21:S4          & 2MASX     
& 14 08 33.28 & -21 36 07.18 & 09.21$\pm$0.08 & 1.18 & -20.72$\pm$0.08 & -13.35$\pm$0.12 & 20.92$\pm$0.02 &[S0]&9137\\
J1408-21:S5*	     & -			      & 14 08 39.82 & -21 38 14.30 & 00.58$\pm$0.58  & --   & -13.31$\pm$0.52 & -15.16$\pm$0.06 & 22.89$\pm$2.24 &[cD]&8788\\ 
J1408-21:S6*	     & -			      & 14 08 52.84 & -21 42 07.20 & 01.73$\pm$0.58  & --   & -17.12$\pm$0.07  & -14.30$\pm$0.48 & 16.70$\pm$0.73 &[dS]&8682\\ 
J1956-50:S1          & IC4909                         & 19 56 45.51 & -50 03 20.29 & 15.70$\pm$0.19 & 2.42 & -21.68$\pm$0.19 & -12.38$\pm$0.16 & 20.70$\pm$0.03 &SA:(rs:)bc&7634\\
J1956-50:S2          & 2MASX         
& 19 55 53.21 & -50 02 10.82 & 08.03$\pm$0.11 & 1.29 & -20.32$\pm$0.11 & -12.69$\pm$0.08 & 20.81$\pm$0.02 &[S/Irr]&7039\\
J1956-50:S3          & -                       & 19 56 08.20 & -50 02 21.56 & 01.72$\pm$0.22 & 1.01 & -16.49$\pm$0.22 & -13.67$\pm$0.05 & 21.59$\pm$0.11 &[BCD]&6400\\
J1956-50:S4*	     & -			      & 19 55 45.92 & -50 06 15.50 & 01.74$\pm$0.74 & --   & -15.15$\pm$0.23 & -14.60$\pm$0.10 & 19.23$\pm$0.96 &[cD]&7497\\ 
J2027-51:S1          & ESO234-G032                    & 20 28 06.39 & -51 41 29.83 & 12.49$\pm$0.29 & 1.89 & -21.15$\pm$0.29 & -12.03$\pm$0.06 & 20.34$\pm$0.03 &(R')SB(s)bc&5805\\
J2027-51:S2          & ESO234-G028                    & 20 27 31.97 & -51 39 20.81 & 16.13$\pm$0.17 & 1.66 & -21.31$\pm$0.17 & -12.13$\pm$0.08 & 20.71$\pm$0.02 &SAB(s)bc pec&5805\\ 
J2027-51:S3          & MRSS
& 20 27 48.52 & -51 44 19.35 & 05.57$\pm$0.12 & 1.59 & -18.93$\pm$0.12 & -13.20$\pm$0.11 & 21.07$\pm$0.03 &[dIrr]&5805\\
J2027-51:S4          & -                       & 20 27 54.64 & -51 38 04.52 & 02.05$\pm$0.15 & 1.15 & -17.20$\pm$0.15 & -13.71$\pm$0.14 & 20.74$\pm$0.04 &[cD]&5988\\
J2318-42a:S1         & NGC7582                        & 23 18 23.44 & -42 22 11.94 & 51.63$\pm$0.26 & 1.98 & -21.60$\pm$0.26 & -11.02$\pm$0.07 & 20.04$\pm$0.02 &(R')SB(s)ab Sy2&1436\\
J2318-42a:S2         & NGC7590                        & 23 18 54.78 & -42 14 18.94 & 26.70$\pm$0.05 & 2.15 & -20.71$\pm$0.05 & -11.12$\pm$0.03 & 19.64$\pm$0.02 &S(r?)bc Sy2&1457\\
J2318-42a:S3***         & NGC7599                        & 23 19 21.14 & -42 15 24.6 & 38.99$\pm$0.29 & 2.03 & -19.76$\pm$0.29 & -11.87$\pm$0.06 & 21.53$\pm$0.02 &SB(s)c&1753\\ 
J2318-42a:S4*	     & APMUKS
  & 23 18 50.44 & -42 23 50.30 & 00.49$\pm$0.41& --   & -11.45$\pm$0.17  & -14.77$\pm$0.27 & 20.72$\pm$1.82 &[LSBD]&1661\\ 
\hline
\end{longtable}
\end{landscape}
\twocolumn
\appendix
\section{Notes on individual Choir groups}

By ``member'' we refer to objects with apparent H$\alpha$ emission in the filter used for the SINGG images. We note that these are likely groups; spectroscopic redshifts are needed to confirm membership, especially for the small, faint galaxies. We also searched larger 40-arcminute photographic survey images\footnote{Digitised Sky Survey images in the blue ($B_J$) band from the Canadian Astronomy Data Centre.} centred on the brightest member of each group (named ``S1'') to check for any bright galaxies that could be group members.

{\bf HIPASS\ J0205-55}: The field  \HIPASS\ J0205-55 covers two sources: \HIPASS\  J0205$-$55a at $V_{\rm hel} = 6524\, {\rm km\, s^{-1}}$ and  \HIPASS\ J0205$-$55b at $V_{\rm hel} = 5964\, {\rm km\, s^{-1}}$ \citep{Meyer2004}.  We note that  \HIPASS\ J0205$-$55a is included in the SINGG sample selection while  \HIPASS\ J0205$-$55b is not \citep{Meurer2006},  Our observations show  a total of 9 galaxies in this rich field: 4 giant spirals and 5 dwarfs of varying sizes. The smallest (S8, S9) are almost in the ELdot category.  The galaxies S1, S2, S3, S4, and S6 have published velocities of 6528, 5927, 6131, 5864, and 5756 km s$^{-1}$ respectively \citep{daCosta1991}.  Hence S1 is associated with  \HIPASS\ J0205$-$55a; S2, S4 and S6 are associated with  \HIPASS\ J0205$-$55b; while S3 is at an intermediate velocity. The existence of galaxies at velocities between the a and b components suggest that the two component systems are merging. The extended optical image of this group reveals one additional large galaxy, ESO153-G020 (velocity 5197 km s$^{-1}$) associated with  \HIPASS\ J0205$-$55b \citep{Doyle2005}.

{\bf HIPASS\ J0209$-$10}: The galaxies of this group show strong signs of interactions,
all being classified ``pec'' and most having extensive extraplanar gas
in the H$\alpha$ images. The group appears in several group
catalogues, most notably it is Hickson Compact Group 16. We found no
new H$\alpha$-emitting galaxies compared to \citet{Meurer2006}
which has a more detailed description of the members in its Appendix
B. (There is a fainter galaxy 1.5' to the NE of S3 = NGC0835, SDSS
J020928.18-100653.6 but it is a background object, velocity = 25706 km s$^{-1}$.)
The extended optical image of this group reveals one additional large
galaxy, NGC 0848 (velocity 3989 km s$^{-1}$) also likely to be
associated with the group \citep{Garcia1993}.

{\bf HIPASS\ J0258$-$74}: A typical small group with three spirals and one tiny dwarf
irregular galaxy. 

{\bf HIPASS\ J0400$-$52}: {Part of} an extensive {cluster (Abell 3193)} with a total of 9 members identified: 4
spirals and 5 dwarfs of varying sizes; two of these are very small
companions to the giant S4 and S6 galaxies. The extended optical image
of this group reveals two additional large galaxies, NGC 1506 (10271
km s$^{-1}$) and ESO156-G031 (10467 km s$^{-1}$) at 10 and 15
arcminutes from the central galaxy S1 respectively. These are both
classified as S0 galaxies, so although associated with the group they
are unlikely to contain large amounts of HI.

{\bf HIPASS\ J0443$-$05}: An extended group of three large spirals, two with
companions. The line emission of S5, an apparent companion to S1,
is weak and needs to be confirmed.

{\bf HIPASS\ J1026$-$19}: This group is dominated by a single face-on giant spiral (S1) which
is connected to S2 by a tidal tail. The 4 other members are small and
well-separated, notably S3 which is on the very edge of the image. 

{\bf HIPASS\ J1051$-$17}: This extensive group has 9 members distributed over much
of the image. The galaxy S9 is notable for being an apparent dE,N
galaxy with weak nuclear H$\alpha$ emission. The extended images
reveals one additional large Sa galaxy, MCG-03-28-016 (6220 km
s$^{-1}$, 9 arcminutes from S1) which may possibly be associated with
the group (5491 km s$^{-1}$).

{\bf HIPASS\ J1059$-$09}: This group features a strongly-interacting galaxy pair (S1
and S3) as well as several other spirals. The two newly-measured galaxies are S9, a small, lopsided dwarf with one \HII\ region, and S10, an edge-on disk galaxy with faint apparent residual H$\alpha$ in the central region as well as weak, very low surface brightness H$\alpha$  along the NW minor axis. S10 is a confirmed group member (2MASX
J10590262$-$0953197 at velocity 8229 km s$^{-1}$) and there are signs of interaction between it and S8, a
possible low-surface brightness group member. The extended image
reveals a bright galaxy, MCG-01-28-020, at 15 arcminutes from S1 but
its velocity (11779 km s$^{-1}$) makes it a background object.

{\bf HIPASS\ J1159$-$19}: This compact group of 4 galaxies features a nearly face-on
late-type spiral with bright H$\alpha$ emission, and 3 dwarfs to the S
and SE. {The field is also known as Arp 022 and is near to the well-known Antennae group, Arp 244.}

{\bf HIPASS\ J1250$-$20}: This is a typical group with two large spirals and 3 dwarf
companions, but we also note the detection of two very compact
H$\alpha$ emitters (S6 and S7) that may be on a tidal tail extending
from S1. These are strong candidates for tidal dwarf galaxies in
formation.

{\bf HIPASS\ J1403$-$06}: This small group (4 members) is dominated by two
strongly-interacting spirals {catalogued as Arp 271}, and also contains two faint ELdot-like
dwarfs.

{\bf HIPASS\ J1408$-$21}: The central galaxy of this group, S1 shows extended
emission. The arm pointing South to S3 shows possible tidal distortion
in the H$\alpha$ emission. There are two new galaxies in the field: S5 and S6. S5 is barely resolved with a single faint \HII\ region and located to the SW of S3, possibly at the extreme end of the tidal arm extending from S1. S6 appears to have weak residual H$\alpha$ in the nuclear region of a small, high-inclination disk, but may be due to bad continuum subtraction in a background galaxy.
 The extended image reveals a bright galaxy, ESO 578-G030, 11
arcmin from S1 but its velocity (10891 km s$^{-1}$) makes it a
background object.

{\bf HIPASS\ J1956$-$50}: This group consists of a large spiral, S1 to the East, a
late type spiral or irregular, S2, to the West, a nearly ELdot-like
blue compact dwarf, S3, projected between them and a new, faint compact dwarf S4 near the W edge of the frame which is difficult to spot due to nearby bad columns in the data. 
{The large velocity spread of these objects (see Table B1) indicates that group membership needs to be confirmed for this group.}

 {\bf HIPASS\ J2027$-$51}: This group contains two large distorted spirals, S1 and S2, a dwarf
irregular, S3, and compact near ELdot dwarf, S4.  The data are relatively noisy,
so may contain faint undetected members in addition to the 4 listed.

 {\bf HIPASS\ J2318$-$42a}: This nearby (1603 km s$^{-1}$) group consists of four large spiral galaxies: NGC 7582, NGC 7590, NGC 7599, plus NGC 7552 which is not visible in the fields of our optical images. The group is known as the ``Grus Quartet'' \citep[see][]{Koribalski2004}. We have identified one very faint additional group member in our H$\alpha$ imaging, denoted S4 in our table: this is one of the faintest group dwarf galaxies in our sample, but follow-up observations have confirmed that it is a group member (Sweet et al., in prep.).

\section{Images}

\begin{figure}
\centerline{
\includegraphics[width=\linewidth]{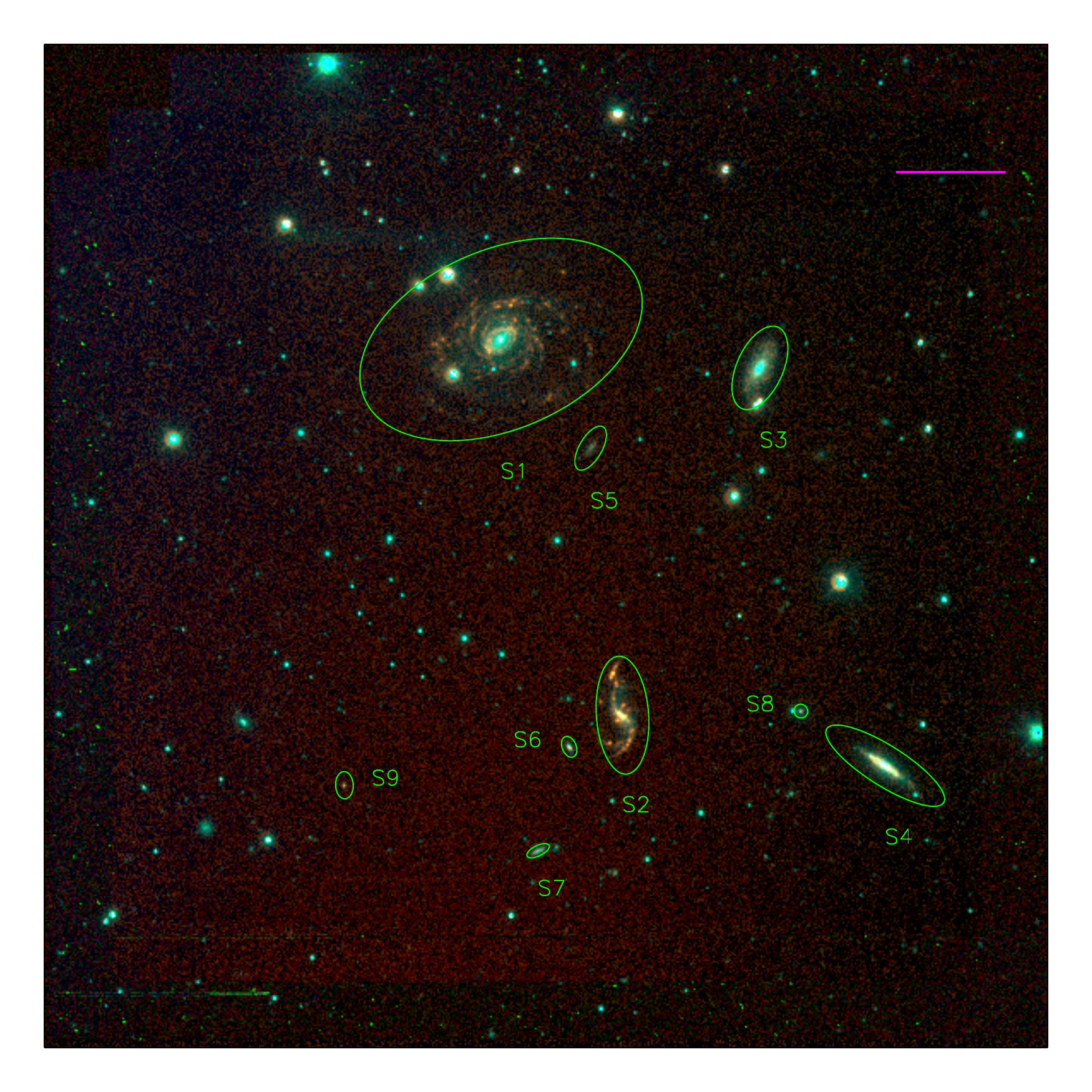} 
}
\caption{Choir group at \HIPASS\ J0205-55. Colours are assigned as follows: R is displayed in the blue channel, the narrow-band H$\alpha$ in the green channel, and the net H$\alpha$ shown in the red channel. ELGs thus appear red. Aperture colours are as follows: green denotes ELGs measured in SINGG, while yellow indicates newly-discovered ELGs. Each image is 15.5 arcminutes on a side. The magenta scale bars indicate 50kpc. North is up and East is left. (Figures~\ref{J0205-55} to~\ref{J2318-42a} make use of this colour scheme, scale and orientation.)
\label{J0205-55}}
\end{figure}

\begin{figure}
\centerline{
\includegraphics[width=\linewidth]{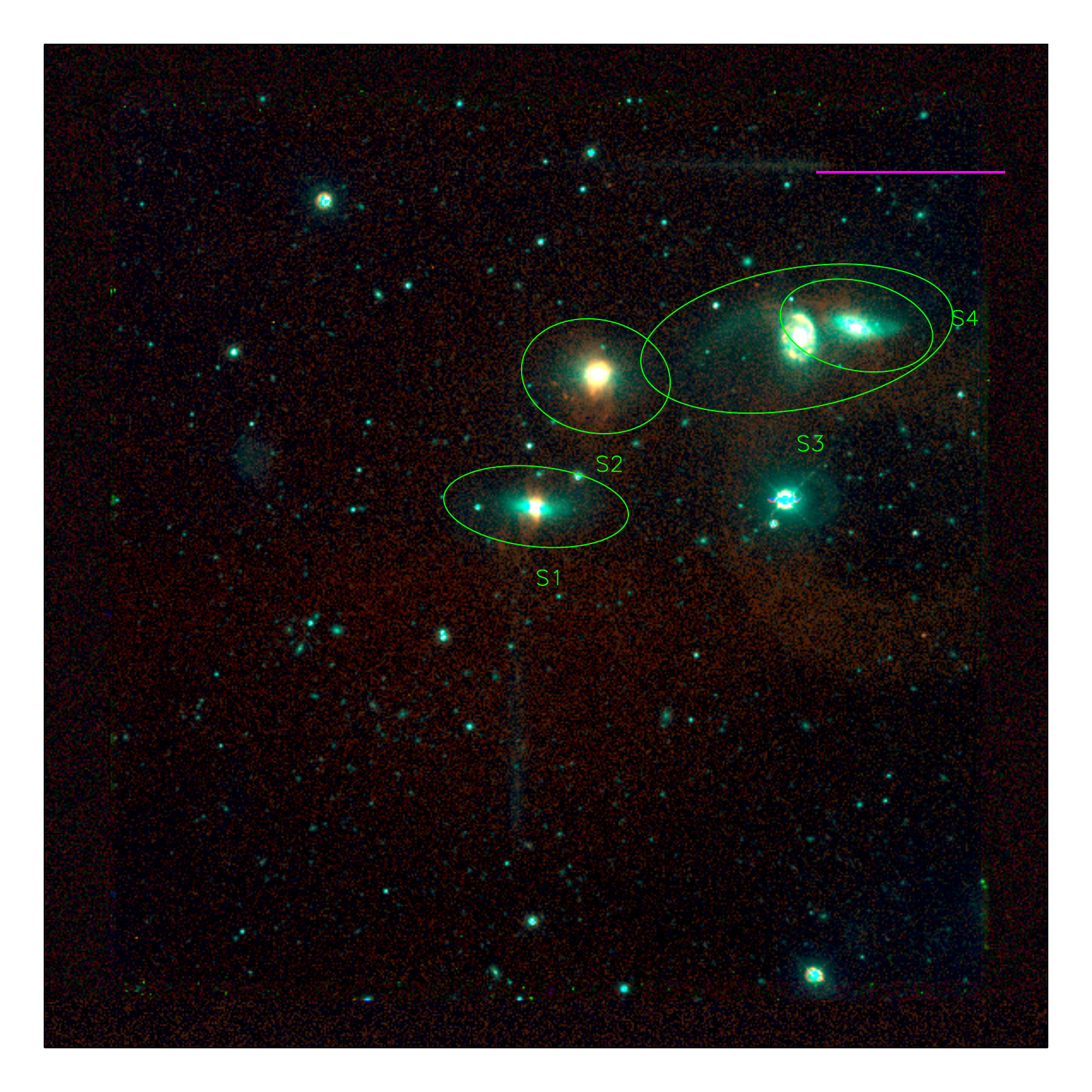}
}
\caption{ \HIPASS\ J0209-10
\label{J0209-10}}
\end{figure}

\begin{figure}
\centerline{
\includegraphics[width=\linewidth]{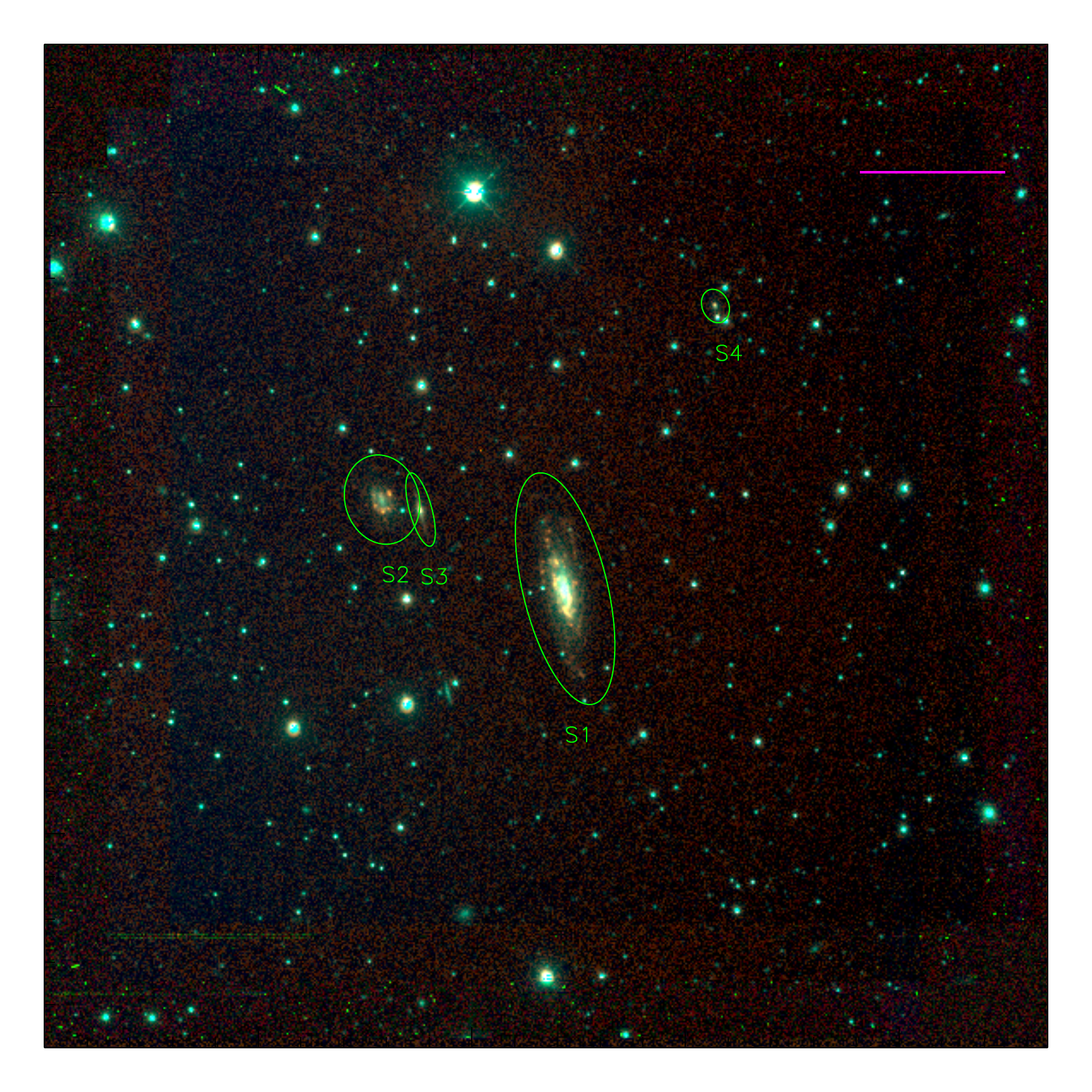}
}
\caption{ \HIPASS\ J0258-74
\label{J0258-74}}
\end{figure}

\begin{figure}
\centerline{
\includegraphics[width=\linewidth]{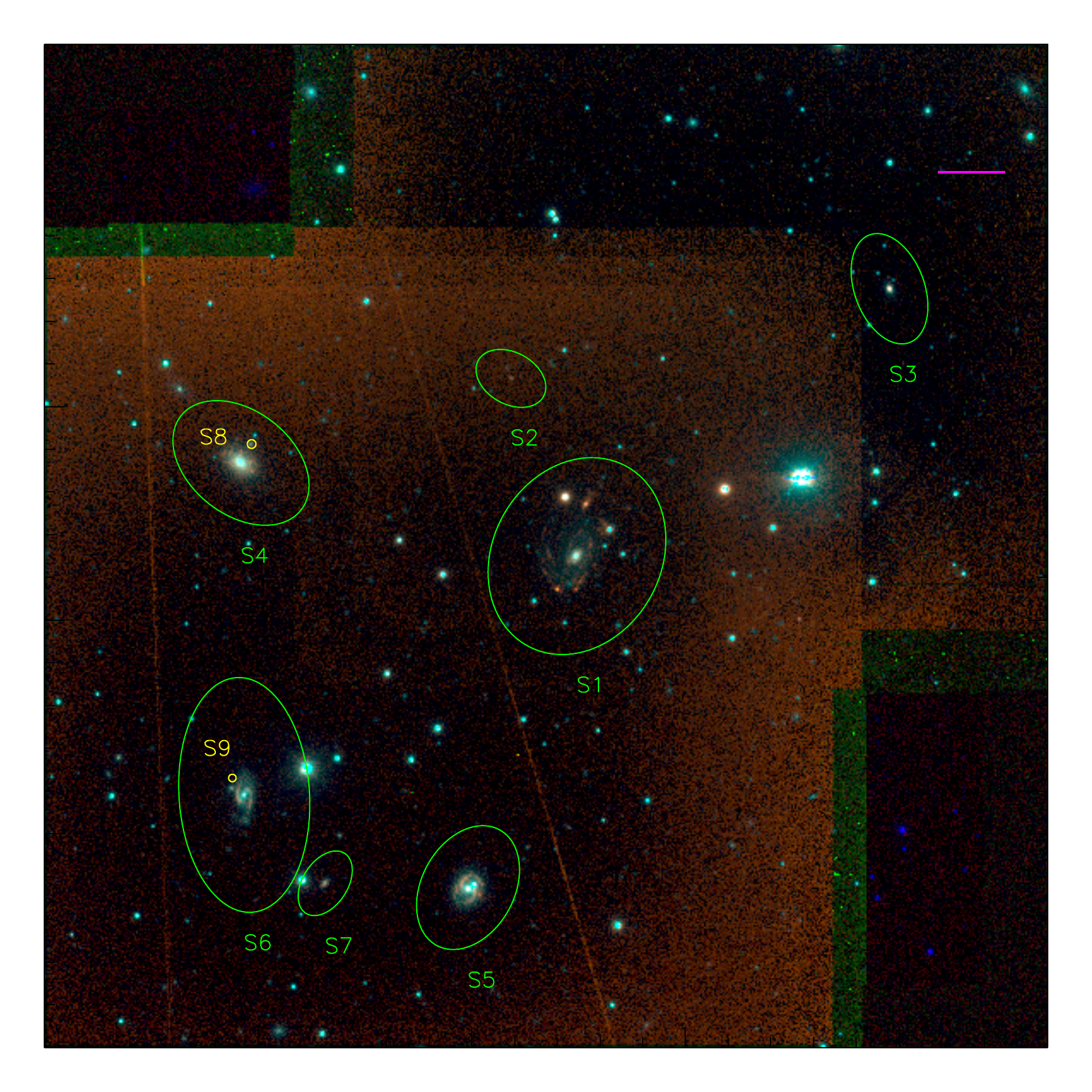}
}
\caption{\HIPASS\ J0400-52
\label{J0400-52}}
\end{figure}

\begin{figure}
\centerline{
\includegraphics[width=\linewidth]{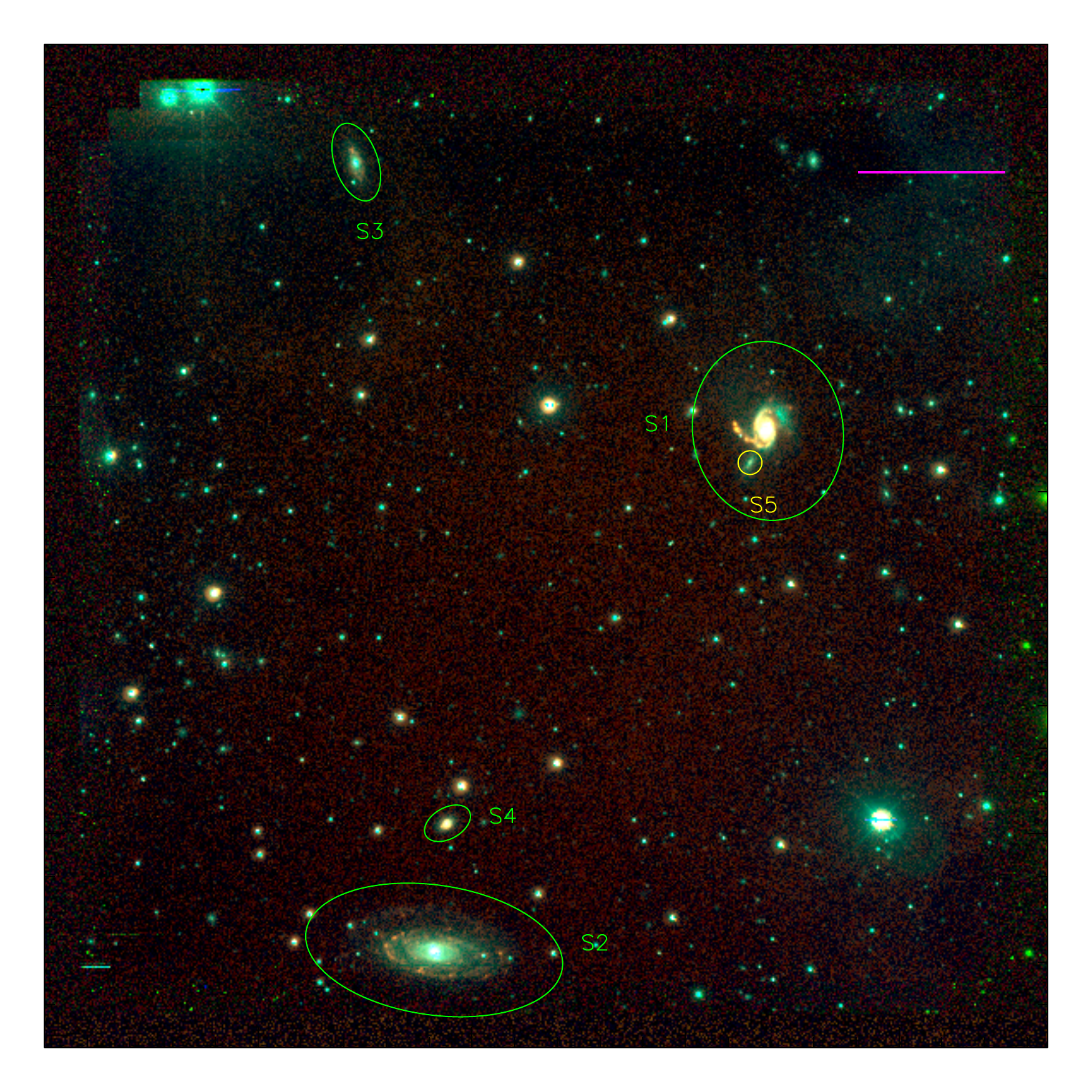}
}
\caption{ \HIPASS\ J0443-05
\label{J0443-05}}
\end{figure}

\begin{figure}
\centerline{
\includegraphics[width=\linewidth]{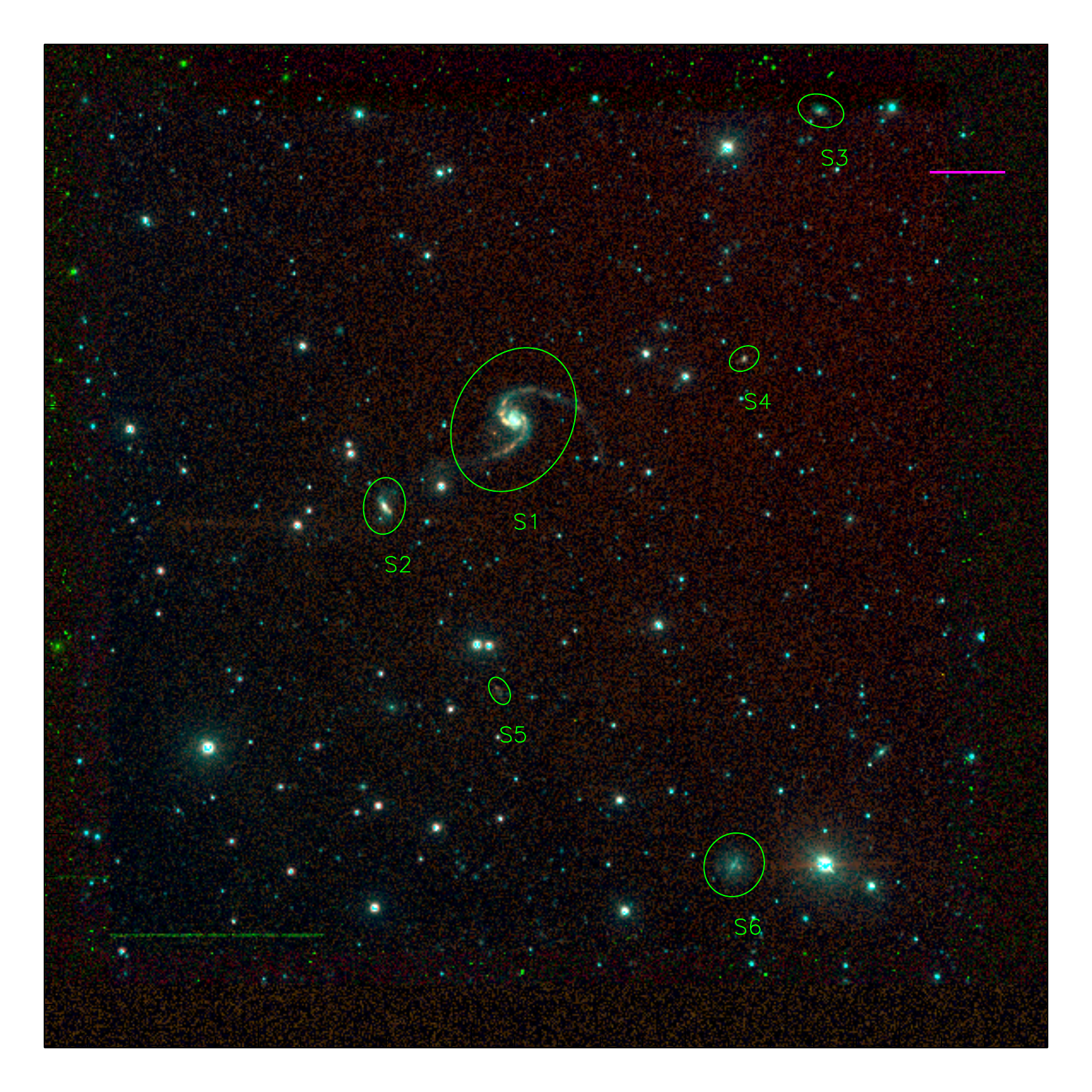}
}
\caption{ \HIPASS\ J1026-19
\label{J1026-19}}
\end{figure}

\begin{figure}
\centerline{
\includegraphics[width=\linewidth]{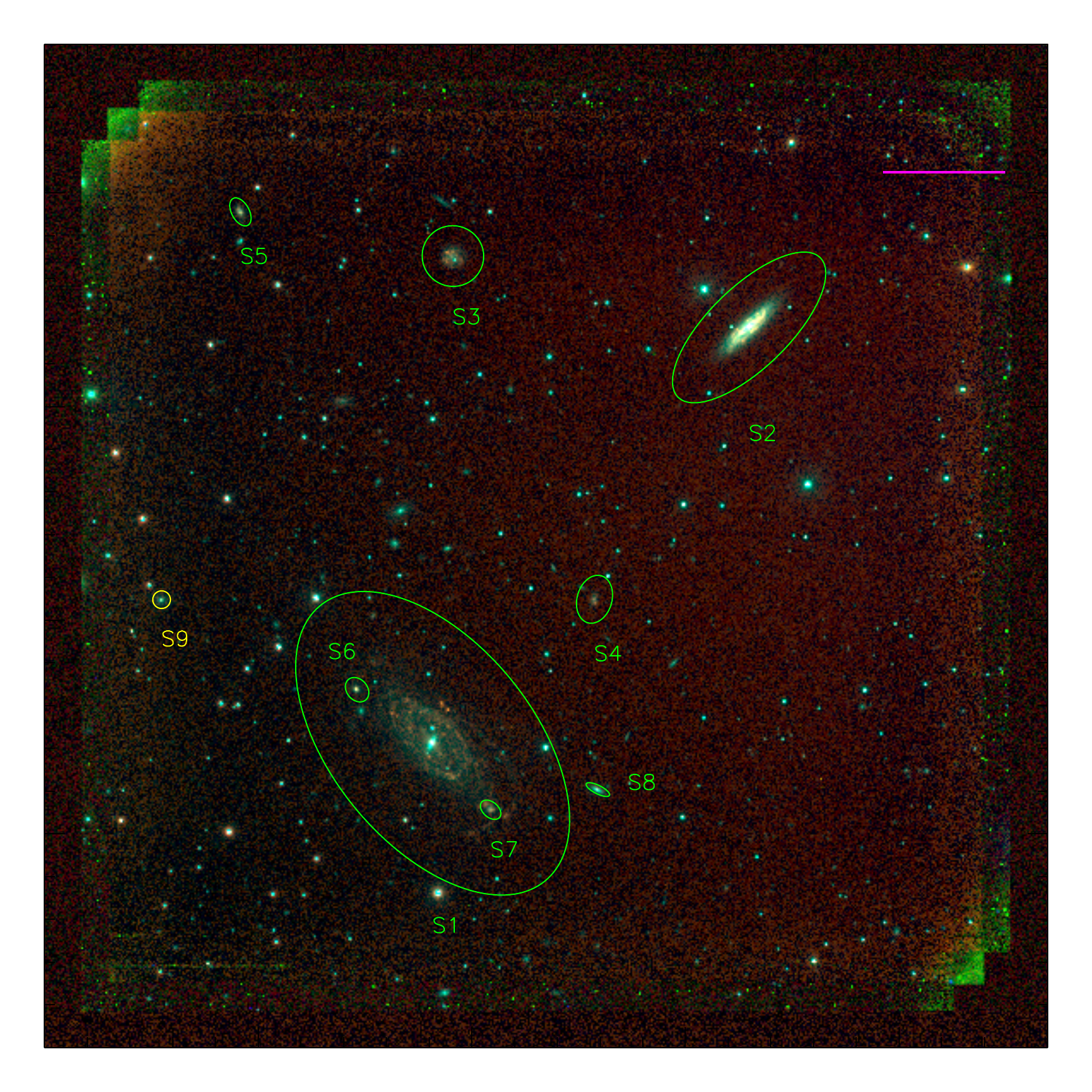}
}
\caption{ \HIPASS\ J1051-17
\label{J1051-17}}
\end{figure}

\begin{figure}
\centerline{
\includegraphics[width=\linewidth]{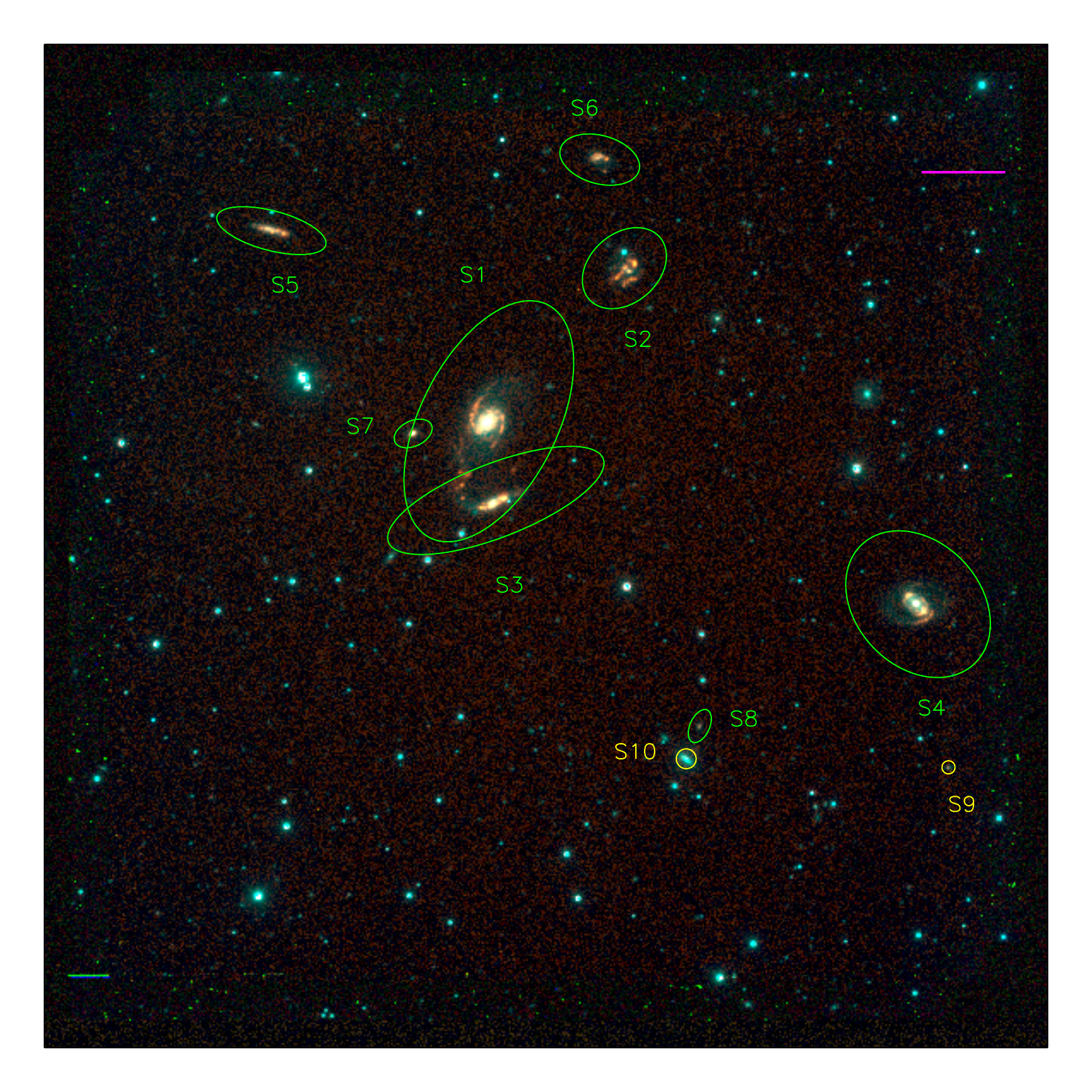}
}
\caption{ \HIPASS\ J1059-09
\label{J1059-09}}
\end{figure}

\begin{figure}
\centerline{
\includegraphics[width=\linewidth]{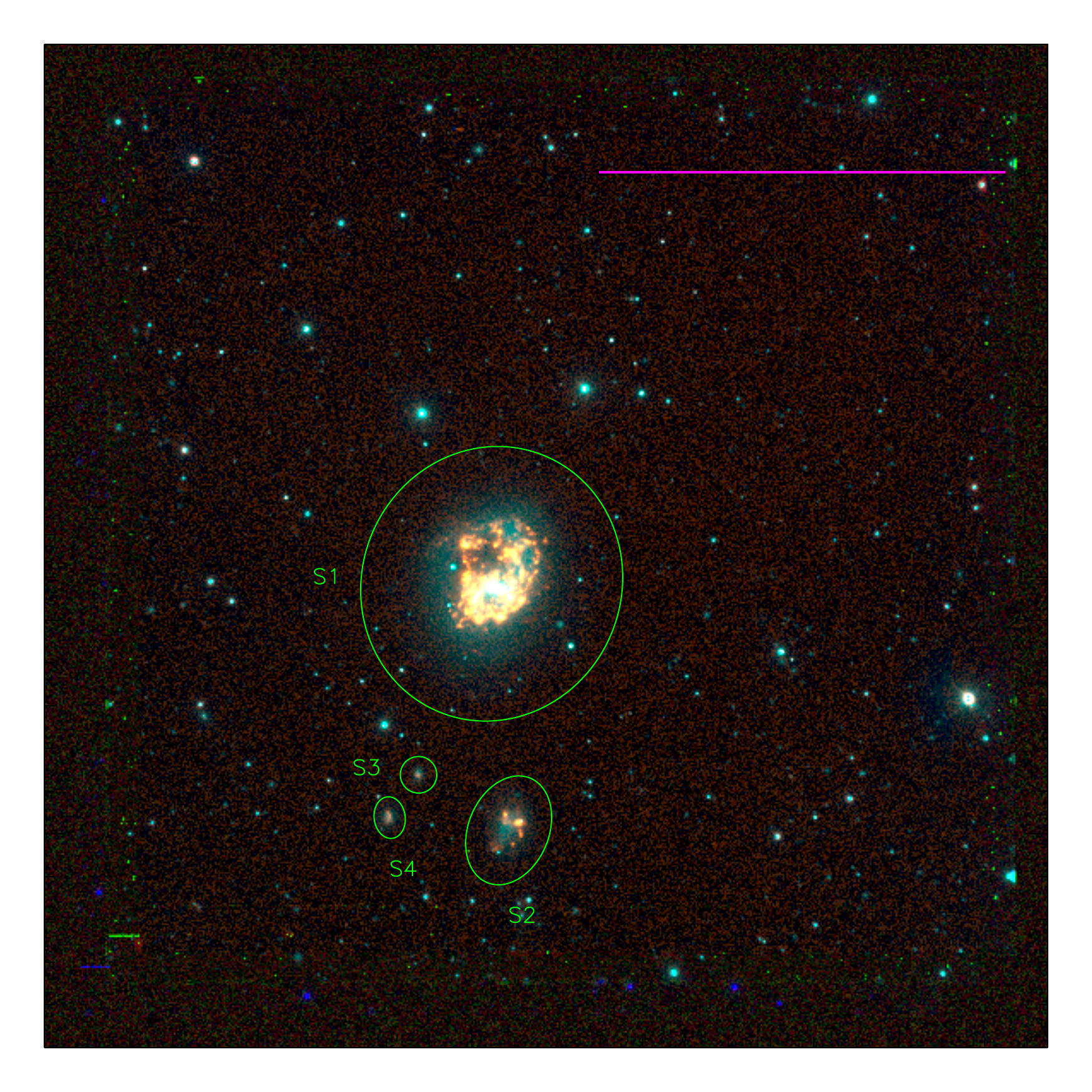}
}
\caption{ \HIPASS\ J1159-19
\label{J1159-19}}
\end{figure}

\begin{figure}
\centerline{
\includegraphics[width=\linewidth]{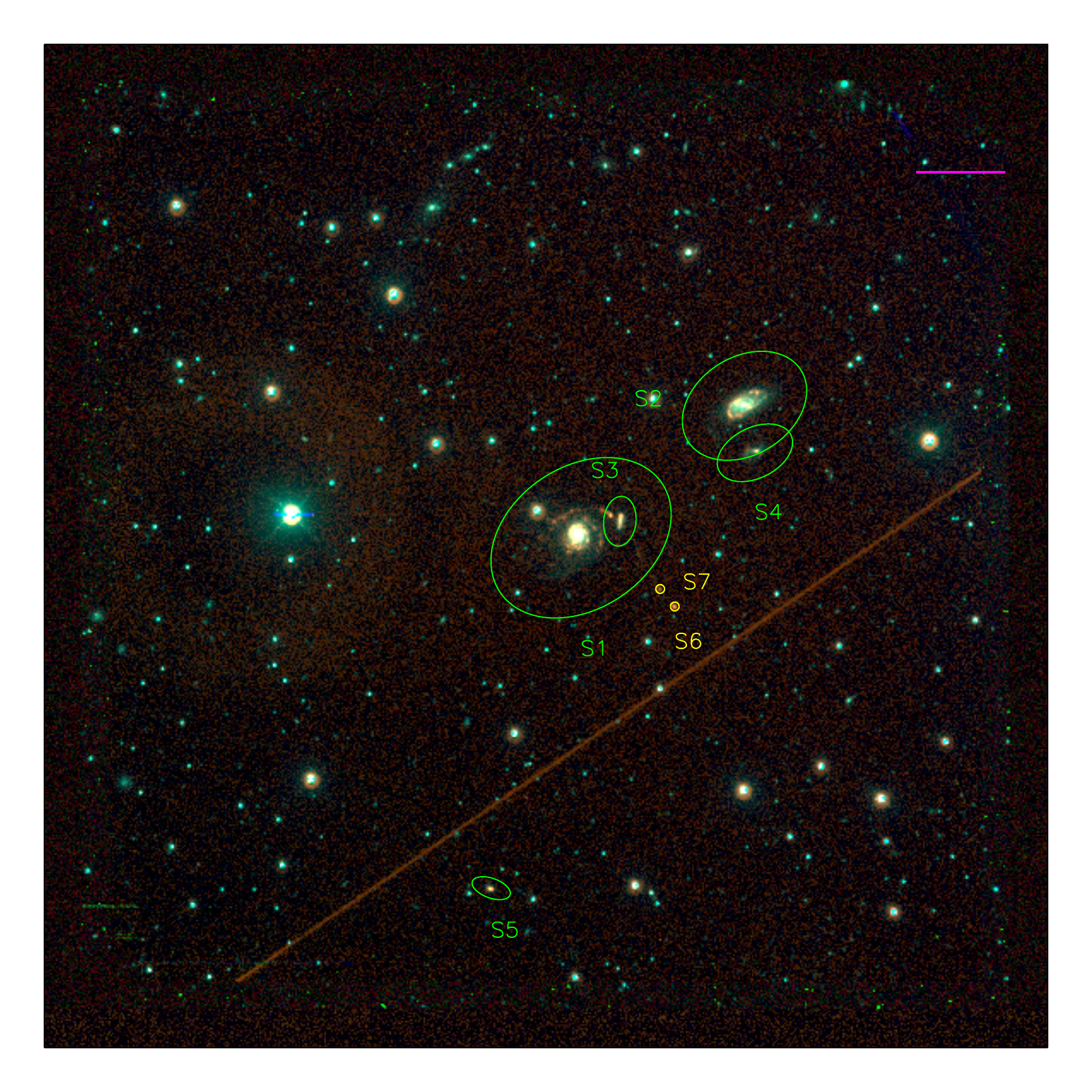}
}
\caption{ \HIPASS\ J1250-20
\label{J1250-20}}
\end{figure}

\begin{figure}
\centerline{
\includegraphics[width=\linewidth]{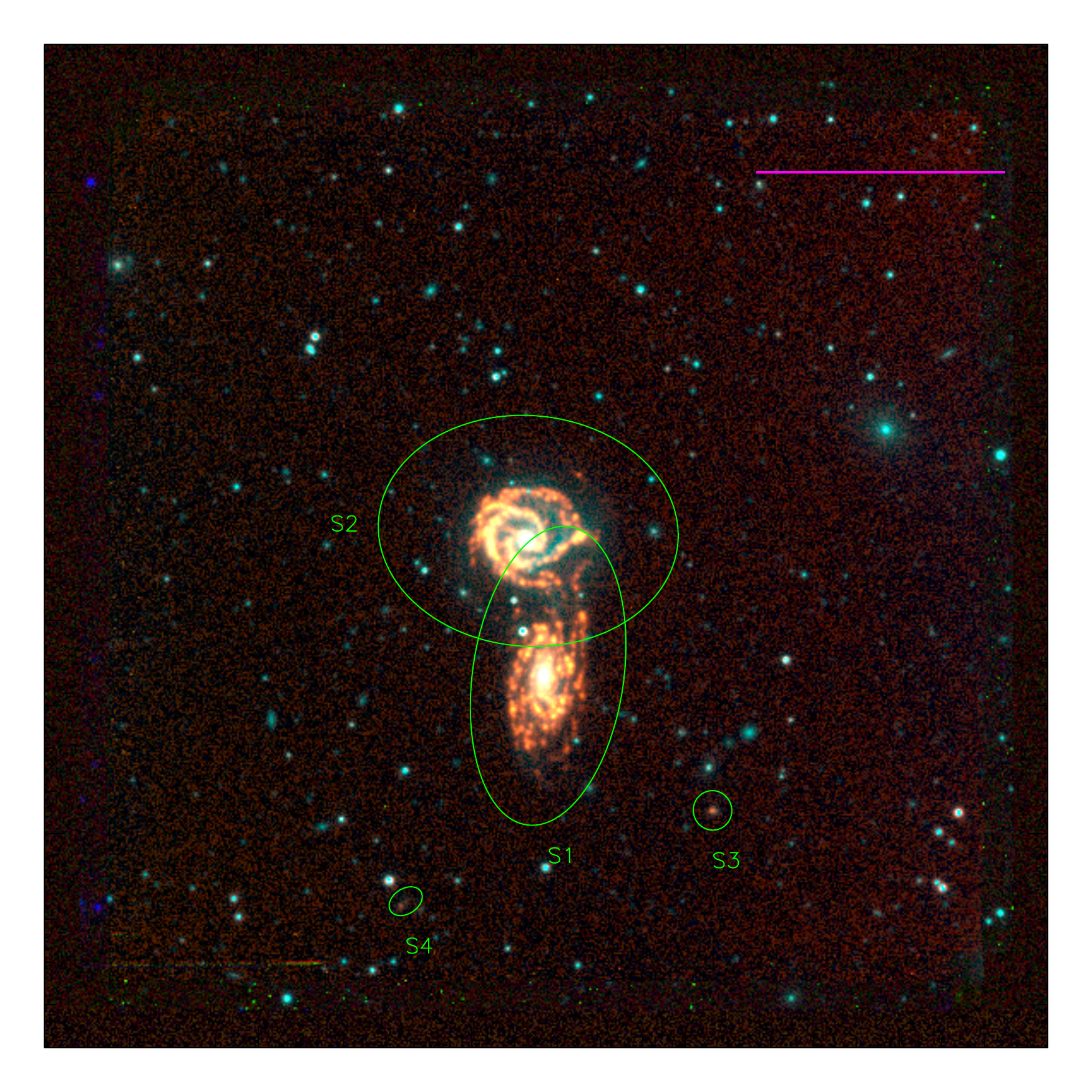}
}
\caption{ \HIPASS\ J1403-06
\label{J1403-06}}
\end{figure}

\begin{figure}
\centerline{
\includegraphics[width=\linewidth]{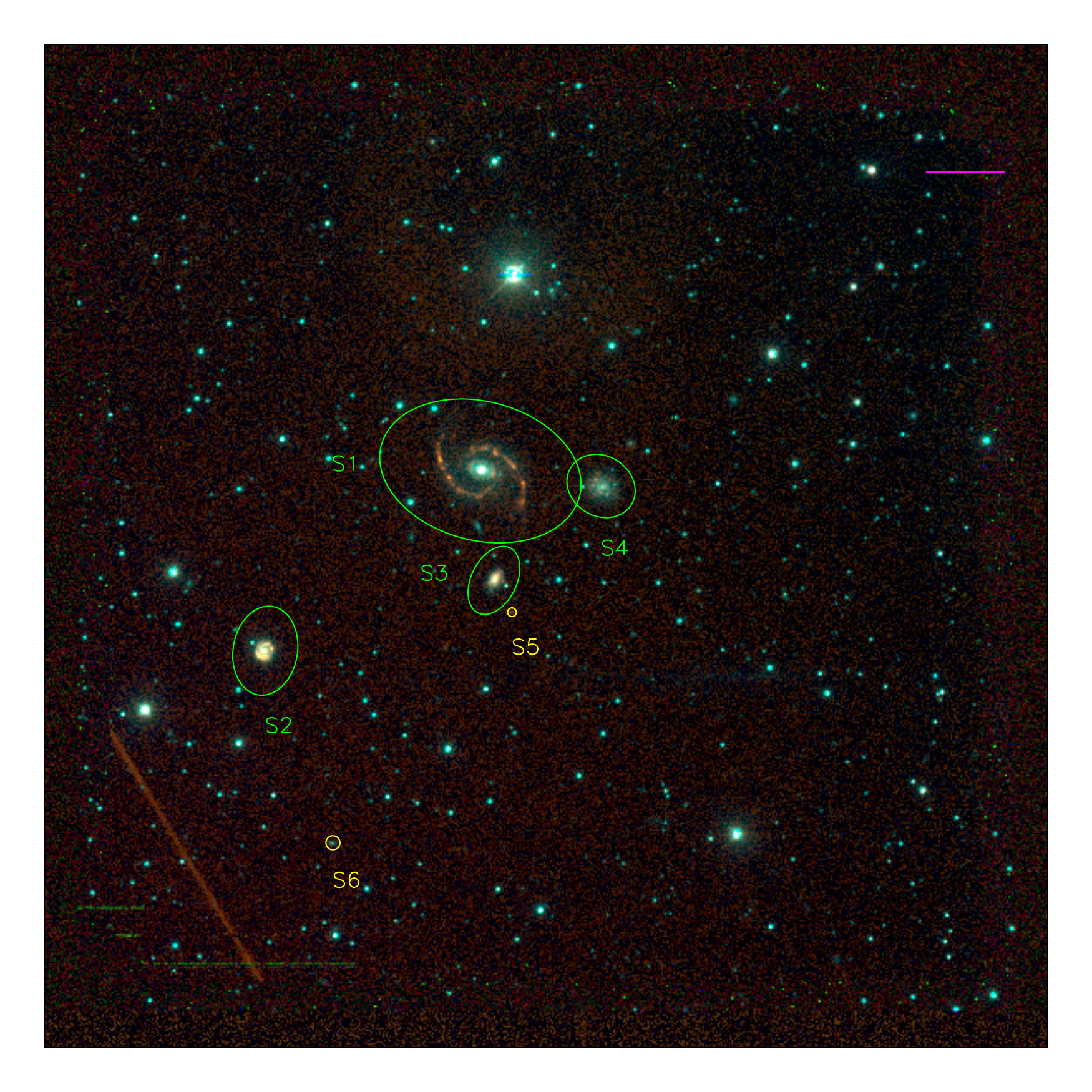}
}
\caption{ \HIPASS\ J1408-21
\label{J1408-21}}
\end{figure}

\begin{figure}
\centerline{
\includegraphics[width=\linewidth]{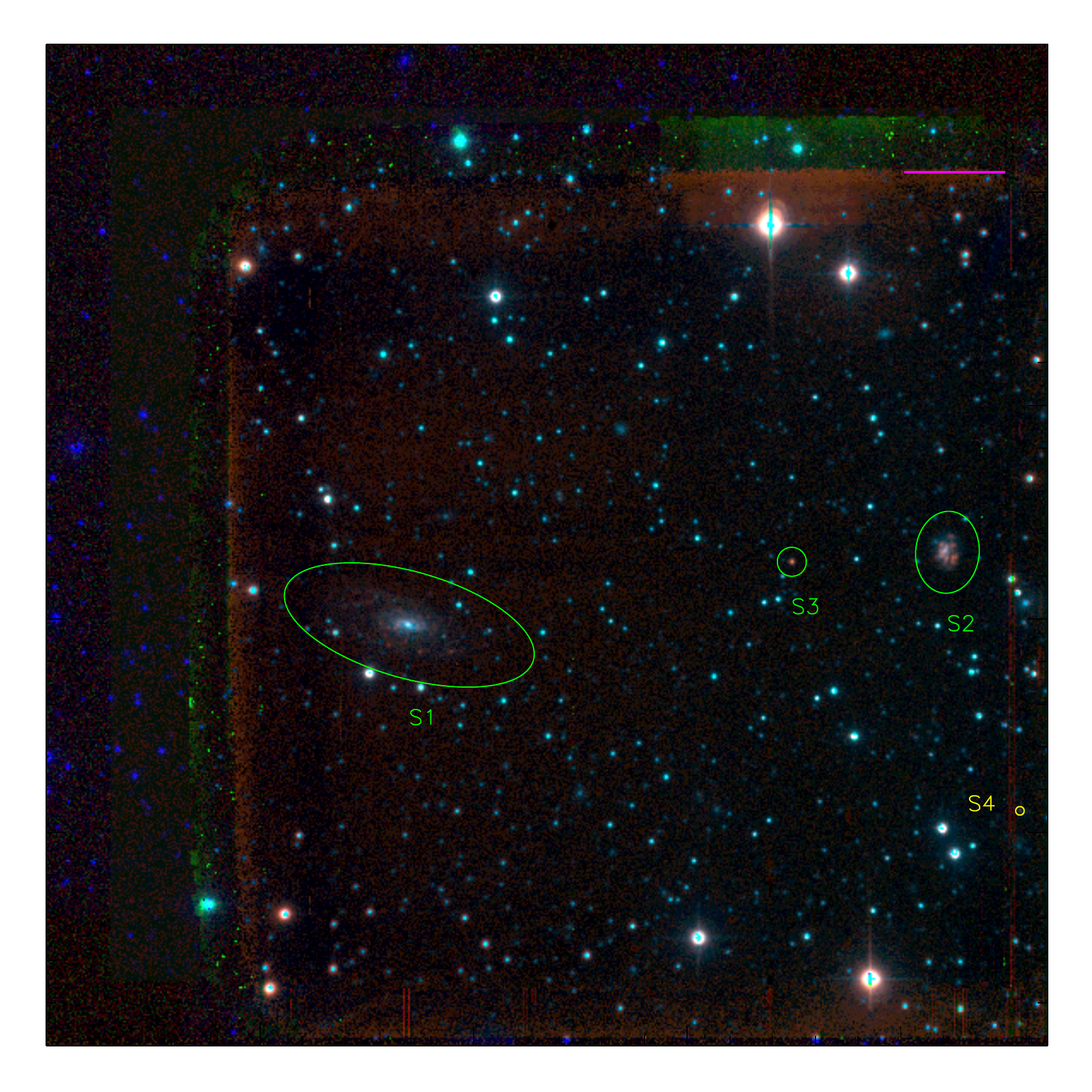}
}
\caption{ \HIPASS\ J1956-50
\label{J1956-50}}
\end{figure}

\begin{figure}
\centerline{
\includegraphics[width=\linewidth]{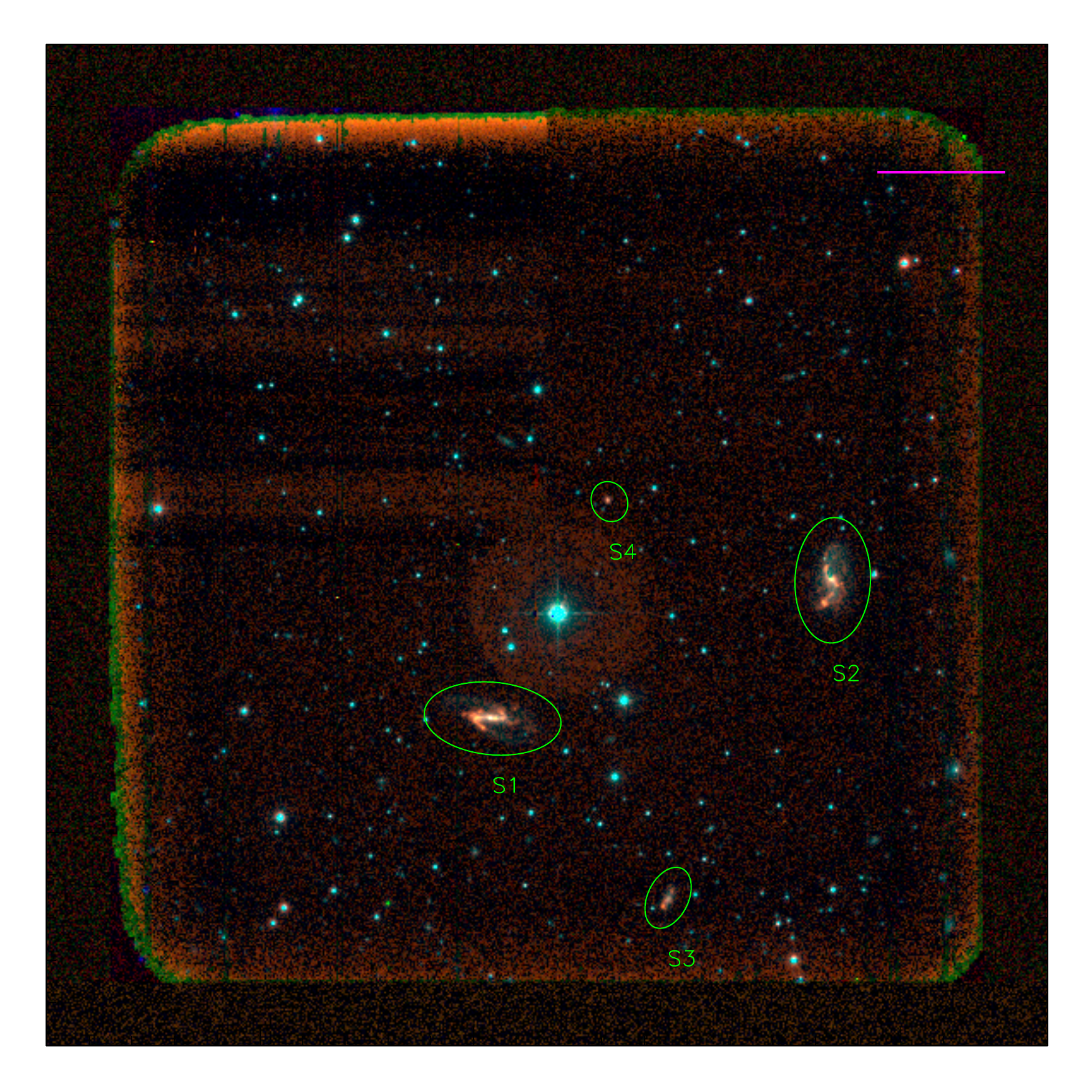}
}
\caption{ \HIPASS\ J2027-51
\label{J2027-51}}
\end{figure}

\begin{figure}
\centerline{
\includegraphics[width=\linewidth]{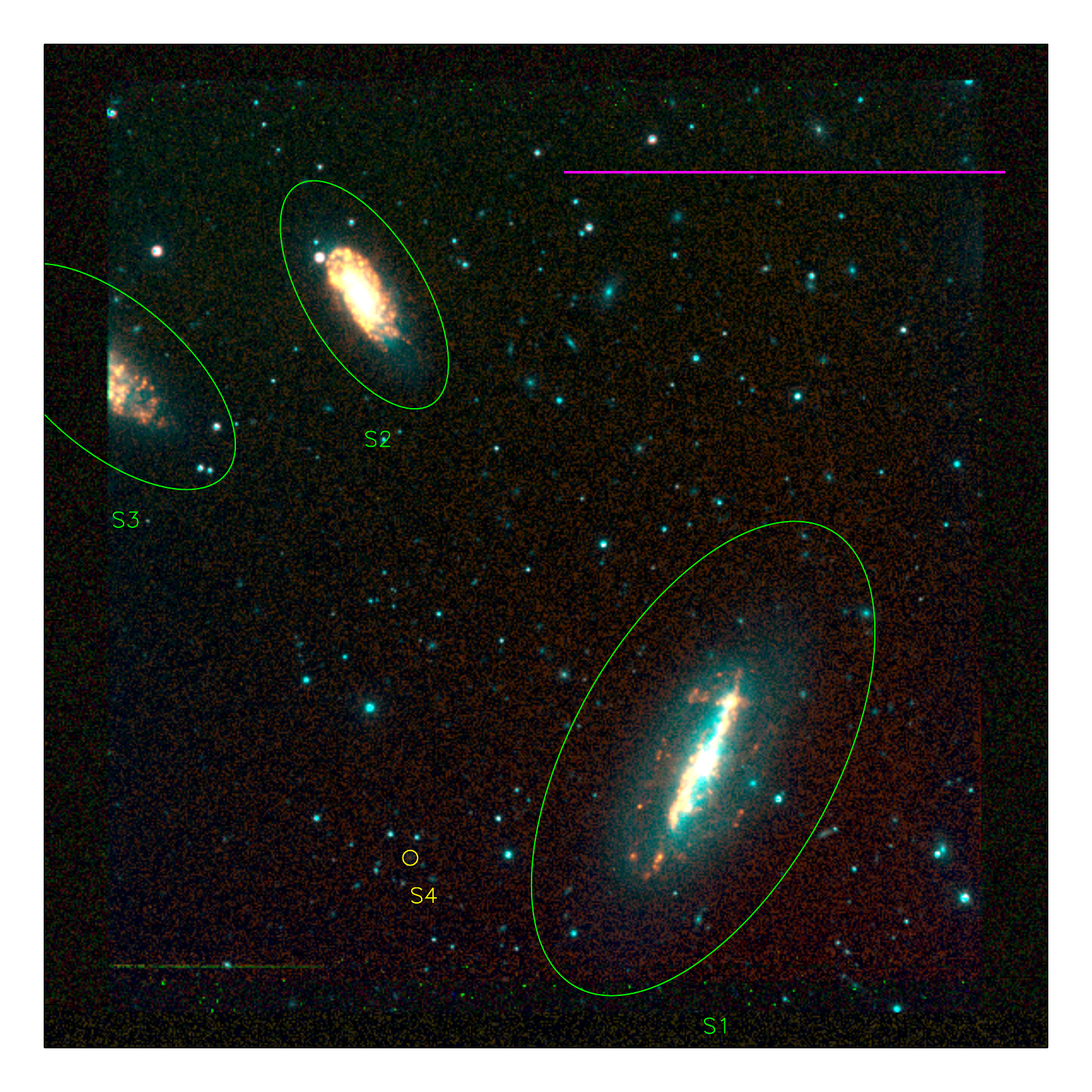}
}
\caption{ \HIPASS\ J2318-42a
\label{J2318-42a}}
\end{figure}

\end{document}